\documentclass{aastex61}

\begin{document}


\title{Modeling the Variability of Active Galactic Nuclei by Infinite Mixture of Ornstein-Uhlenbeck(OU) Processes}

%% Use \author, \affil, plus the \and command to format author and affiliation 
%% information.  If done correctly the peer review system will be able to
%% automatically put the author and affiliation information from the manuscript
%% and save the corresponding author the trouble of entering it by hand.
%%
%% The \affil should be used to document primary affiliations and the
%% \altaffil should be used for secondary affiliations, titles, or email.

%% Authors with the same affiliation can be grouped in a single
\correspondingauthor{Tadafumi Takata}
\email{tadafumi.takata@nao.ac.jp}

\author{Tadafumi Takata}
\affiliation{Astronomy Data Center, 
National Astronomical Observatory of Japan, 
National Institutes of Natural Science(NINS)
Osawa 2-21-1 
Tokyo, 181-8588, JAPAN}
\affiliation{The Graduate University for Advanced Study (SOKENDAI)}

\author{Yusuke Mukuta}
\affiliation{Graduate School of Information Science and Technology, The University of Tokyo}
%% Use the \and command so offset the last author.
%\and
\author{Yoshikiko Mizumoto}
\affil{National Astronomical Observatory of Japan, 
National Institutes of Natural Science(NINS)}


\begin{abstract}
We develop an infinite mixture model of Ornstein-Uhlenbeck(OU) processes for describing the optical variability 
of QSOs based on treating the variability as a stochastic process. This enables us to 
get the parameters of the power spectral densities(PSDs) on their brightness variations by providing 
more flexible description of PSDs than the models based on single OU process(damped random walk). 
We apply this model to 67,507 variable objects extracted from SDSS Stripe82 photometric data 
and succeed in showing very high precision in identifying QSOs ($\sim$99\% levels in completeness 
and purity) among variable objects based only on their variability, by investigating on 
9,855 spectroscopically confirmed objects(7,714 QSOs and 2,141 stars) in the data of SDSS Data Release 12(DR12), 
with sufficient and accurate multiple measurements of their brightness. 
By comparing our results with the values based on other models that are used in previous research, 
it is revealed that our model can be used as the most effective method for selecting QSOs from 
variable object catalog, especially regarding completeness and purity. 
The main reason of improved identification rates are the ability of our model to separate clearly 
QSOs and stars, especially on the small fraction of QSOs with variabilities which can be described 
better than simple damped random walk model. 
\end{abstract}

\keywords{Active Galactic Nuclei(AGN) --- Quasi Stellar Objects(QSOs) --- 
optical variability --- stochastic process --- Ornstein-Uhlenbeck(OU) Process --- classification}



\section{Introduction} \label{sec:intro}
One of the remarkable properties of active galactic nuclei (AGNs), frequently 
represented by Quasi Stellar Objects(QSOs), is their variability seen in a wide 
range of the electromagnetic wave region.
Just after the discovery of AGNs\citep{schmidt1963,mattews1963}, 
their variability became well known\citep{greenstein1963,smith1963} and 
studied by many authors writing about theory and observational data analysis. 
Since \citet{sesar2007} demonstrated that at least 90\% of QSOs show the signature of variability 
with rms more than 0.03 mag, it is believed that most QSO/AGN populations have a feature of variability.   
As AGNs are among the most luminous celestial objects in the universe, their variability 
with an order of 10\% of their total light in various time scales 
(less than one hour to many years)\citep{gaskell2003,uttley2014}, are one of the largest fluctuations in energy, 
and discovering their originating physical processes is an attractive quest. 
Because the optical continuum radiation is believed to be predominantly coming from the accretion disk, it is 
straightforward to consider that some processes intrinsic to the disk are the origin of the brightness fluctuations. 
There is such a lot of theoretical work discussing the origin of the variability, 
in the short and long term, based on the change of global accretion rate(for example, \citet{pereyra2006}), 
disk inhomogeneities propagating inward(for example, \citet{dexter2011}), and these fluctuations may arise from 
thermal or magneto-rotational instabilities in a turbulent accretion flow\citep{hirose2009,jiang2013}, 
though it is still unknown what physical processes control their variability. 
\par
On the data analysis side, there is still a lot of research describing their behaviors 
by using structure functions(SFs) and power spectral densities(PSDs). 
In recent years, there has been much progress in the modeling of their  
variability (light curves) by means of massive data with long duration 
and/or dense cadences. There are many extensive studies on QSO/AGN variability, 
as a powerful tool for QSO selection using legacy and/or newly acquired data, and also as a probe 
for physical models of AGNs
\citep{kelly2009,kelly2011,kelly2013,kozlowski2010a,kozlowski2011,kozlowski2012,
kozlowski2016b,macleod2010,macleod2011,macleod2012,schmidt2010,schmidt2012,
Palanque-Delabrouille2012,butler2011,kim2011,ruan2012,zuo2012,andrae2013,
zu2013,morganson2014,graham2014,decicco2015,falocco2015,cartier2015,caplar2017}. 
\par
A model with particularly successful results is the damped random walk (DRW) model. 
This model was first introduced by \citet{press1992} and \citet{rybicki1992}, and 
the fast computational implementation was described by \citet{rybicki1995}, for inferring 
the time lag of variability among multiply imaged gravitationally lensed QSOs. 
With the detailed analysis by \citet{kelly2009}, \citet{kozlowski2010a} and \citet{macleod2010} on various 
time series data, it is established that a DRW model can statistically explain the observed light curves of QSOs 
at the enough fidelity level(0.01 to 0.02 mag). 
Although the DRW model is relatively simple in description, it successfully 
describes the most of QSOs' light curves, leading to very high rates in identifying 
QSOs from the variable sources especially in optical wavelength(for example, 
\citet{kozlowski2010a,macleod2010,macleod2011,choi2014}), although it is revealed to be applicable also to 
mid-infrared data\citep{kozlowski2010b,kozlowski2016b}. 
It should also be noted that the confidence level and completeness of QSO selection will be increased by 
the combination of colors and variability\citep{peters2015}, and/or 
with support by image subtraction technique\citep{choi2014}. 
The model calculation is relatively fast; it needs computational times only scaling to the number of data points
($O(N)$ in case N is the number of data points), by the implementation of SF for DRW model by using 
auto-correlation function\citep{macleod2010, butler2011}, 
although the other implementation using Markov Chain Monte Carlo (MCMC) for reproducing the continuous light curves 
and exponential covariance matrix takes longer in $O(N^2)$\citep{zu2011,zu2013}. 
On the other hand, there are claims that DRW is too simple to describe the 'real' variability of QSOs/AGNs,   
because the possibility of the degeneracies cannot be eliminated\citep{kozlowski2016c}. 
\citet{kasliwal2015a} investigated the optical light curves of 20 QSOs observed with $Kepler$ satellite, and found 
that fewer than half of them can be explained by the DRW model.  
\par
Thus there are some alternatives for describing the variability in Fourier space with PSDs, 
by the model with mixture of Ornstein-Uhlenbech(OU) processes\citep{kelly2011}, 
Slepian wavelet variance\citep{graham2014}, continuous auto regression and moving average(CARMA) 
model\citep{kelly2014,simm2016,kasliwal2017}, broken power law model\citep{zhu2016} and so on. 
\par
The OU mixture model by \citet{kelly2011} is one of the powerful candidates 
of QSO/AGN variability modelings because it is very natural to try to express the fluctuation 
phenomena by the combination of the simple random processes. The method of the OU mixture is based on 
the long-term accumulation of experience learned in many studies of QSO/AGN and 
also of galactic objects' variability.   
As this model treats the variability in Fourier space, we can learn their PSDs, which provide us 
with the information on how they vary. This should be very useful for investigating the physical processes 
which control the light variations. It is also revealed to be possible to apply to some 
wavelength data such as X-ray and optical data\citep{kelly2011}. 
However, the model has some weak points; the first one is their time-consuming calculation, 
and the second one is the arbitrariness in the selection of the number of OU processes to be mixed. 
As suggested in \citet{kelly2011}, the number of mixed OU processes (expressed as $M$ hereafter) 
should be larger than 30 to describe the sufficiently observed AGN's light curves, and the calculation is too 
massive for large sample data, because the calculation time is on the order of $O(MN^2)$, which is too massive 
for those for DRW model calculations. 
\par
We've tried to overcome these weak points by extending the mixture numbers of OU processes to infinite, 
and succeeded in finishing the calculation in much less time than the original model and without the 
arbitrariness in OU mixture. 
By applying our infinite OU mixture model to 67,507 variable sources extracted from 
the photometric data of Stripe 82 in Sloan Digital Sky Survey(SDSS), we confirm the ability of our model 
in selection of QSOs from various variable objects and compare our results with those 
by other models previously suggested as successful. By our model we can show the variation of the PSDs 
of numerous ($\sim$8,000) spectroscopically confirmed QSOs with sufficient multiple photometric measurements 
during $\sim$10 years, and provide some implication for the power of PSDs in classifying QSOs/AGNs into 
subclasses and find some rare type QSOs/AGNs.   
\par
It should be emphasized here that we know it is very difficult to describe the details on intrinsic variability 
and the underlying physical processes of individual QSOs in this study, as the data of SDSS Stripe82 suffer from 
irregular and sparse time sampling, and our model uses approximate PSD model. 
Therefore, our main purpose in this paper is to know where is the limit in distinguishing QSOs from stars 
only using their variability, with superpositions of OU processes, which is more flexible to describe 
variability of QSOs/AGNs than those based on single OU process. 
\par
In section \ref{sec:methodologies}, we will describe the methodology by which we construct our 
model, and describe the data used for our test in section \ref{sec:data}. In section \ref{sec:results}, 
we will show the results by our analysis of Stripe 82 data. 
We show the methods on classification and identification rates based on variability 
in section \ref{sec:id_rate}, discuss the features of our model in section \ref{sec:discussions}, 
and provide the conclusions in section \ref{sec:conclusions}. 
\par

\section{Methodologies} \label{sec:methodologies}
In this section, we describe existing models and our proposed models as
their limit. We first describe the estimation method common to each
model in \ref{subsec:estimate}. We describe OU process in \ref{subsec:drwmodel}, and 
the model with finite mixture of OU process proposed by \citet{kelly2011} in \ref{subsec:mixmodel}. 
Afterward we describe our model with infinite mixture of OU processes in \ref{subsec:ourmodel}, and 
practical implementation and model calculation in \ref{subsec:ourmodel_imple_calc}. 

\subsection{Estimation Method}\label{subsec:estimate}
For estimating the best model parameters, by which the physical process is described, we usually perform 
fitting by using maximum-likelihood or Bayesian techniques. There is an advantage in estimating the 
parameters directly from the light curve by using all of the information in the data. 
It is especially useful to estimate the parameters of power spectrum model as they are not heavily biased by 
measurement errors, irregular sampling, or other windowing effects caused by the finite time span of the 
light curve, such as red noise leak\citep{kelly2011}.
Though the likelihood function may be used to calculate a maximum likelihood estimate, we employ a Bayesian 
approach which calculates the posterior probability distribution of the model parameters, in order to 
reliably estimate the uncertainties of the model parameters. 
The probability distribution of the parameters for the given light curve by the observation is 
\begin{equation}
p(\theta|x_{t_1},...,x_{t_N}) \propto p(x_{t_1},...,x_{t_N}|\theta)p(\theta),
\end{equation}
where $p(\theta)$ and $p(\theta|x_{t_1},...,x_{t_N})$ are the prior and the posterior distributions.  

The likelihood function $p(x_{t_1},...,x_{t_N}|\theta)$ can be written as the following form 
\begin{equation}
p(x_{t_1},...,x_{t_N}|\theta) = \frac{1}{\sqrt{(2\pi)^N|C|}}\times{\rm{exp}}\left(-\frac{{\bf{x}}^T {C}^{-1} {\bf{x}}}{2}\right)
\end{equation}
, where $C$ is covariance and the vector ${\bf{x}}$=$(x_{t_1},x_{t_2},...,x_{t_{N-1}},x_{t_N})$.
Here we denote model parameter as $\theta$ and observed data as $\{x_{t_n}\}_{n=1}^N$, where 
$x_{t_n}$ is a brightness(apparent magnitude) at $t_n$ and $N$ is the number of observations(measurements). 
$N$ is assumed to be about $10 \sim 10^2$ in the data-set. 
Given prior distribution $p(\theta)$ and generation model $p(x_{t_1},...,x_{t_N}|\theta)$, 
we use maximum a posteriori(MAP) estimation as described below 
\begin{eqnarray}
\theta_{\rm MAP} &=& {\rm arg max}_{\theta} p(\theta|x_{t_1},...,x_{t_N}) \\
 &=& {\rm arg max}_{\theta} p(x_{t_1},...,x_{t_N}|\theta) p(\theta),
\end{eqnarray}
to get the model parameter, where $\theta_{\rm MAP}$ means the parameter set giving the maximum of likelihood for the model.
In fact, it is hard to compute the exact $\theta_{\rm MAP}$. Thus, we use MCMC to randomly sample
$\{\theta_i\}_{i=1}^I$ from $p(\theta|x_{t_1},...,x_{t_N})$ and use the median of $\{\theta_i\}_{i=1}^I$ instead, where  
$I$ is a number of samples. In this work, we use Affine Invariant MCMC Ensemble Sampler\citep{foreman-mackey2012} 
and set $I=20,000$ for getting statistically stable results by numerous sampling. 
We tested the cases of $I=5,000$, $10,000$ and $20,000$ and confirmed that there is little difference   
among the inferred parameter values in the most cases, as mentioned by \cite{zu2011}. 
However, as our model will use four parameters, which is twice of those by \cite{zu2011}, and the model 
is more complex, we decided to use $I=20,000$ for our calculation. 

It requires that we can compute $p(x_{t_1},...,x_{t_N}|\theta)$ with
small complexity. All models we handle in this paper are Gaussian process. 
This means that the joint distribution of $\{x_{t_n}\}_{n=1}^N$
is a $N$-dimensional Gaussian distribution with $i,j$-element of covariance 
$C_{ij}={\rm cov}(|t_i - t_j|,\theta)$, where ${\rm cov}$ is a covariance function. 
When ${C}$ is computed, it requires $O(N^3)$ complexity to compute $p(x_{t_1},...,x_{t_N}|\theta)$. 
This does not take much computation because $N$ is small. Thus, our motivation
is to propose a covariance function with high representation ability and
requires small computational complexity.

\subsection{Damped Random Walk(DRW) Model(OU process)}\label{subsec:drwmodel}
The most basic model is OU process. This is a continuous auto-regression
model. The model parameter $\theta$ consists of time constant $\tau$ and noise
magnitude $\zeta$. OU process is a solution of the following stochastic
differential equation:
\begin{equation}\label{eq:ou}
 dX_t=-\frac{1}{\tau}X_tdt+\zeta dW_t,
\end{equation}
where $W_t$ is a white noise.

The covariance function of this model is
\begin{equation}\label{eq:ou_cov}
 \mathcal{R}_{OU} = \frac{\tau\zeta^2}{2}e^{\frac{-|\delta t|}{\tau}}.
\end{equation}

The most important advantage of the OU model is that we need only two parameters for the model($\zeta$ and $\tau$). 
Also, it does not require much computation, as the covariance function is simple. 
Additionally there is a faster implementation for the model computation by using structure function which is 
derived from the auto-correlation function of DRW model\citep{hughes1992,macleod2010}. 

The power spectral density can be derived by Fourier transformation of Equation \ref{eq:ou_cov}, 
\begin{eqnarray}
\mathcal{P}_{OU}(\omega) &=& \frac{1}{2\pi}\int_{-\infty}^{\infty} e^{-i \omega t}\mathcal{R}_{OU} dt\\
              &=& \frac{\zeta^2}{2\pi} \frac{1}{\omega_{0}^2 + \omega^2}, 
\end{eqnarray}
where $\omega_0$ is the reciprocal of damping time scale $\tau$, and it is good approximation of majority of 
QSOs with brightness variation\citep{kelly2009,kozlowski2010a,macleod2010}. 

However, since the model is so simple\citep{kasliwal2015a,kozlowski2016c}, the scope of the method is limited. 
The actual/real process is considered to consist of stochastic process with various scales. 
Thus, it is natural to consider the mixture model for describing the actual and relatively complex processes. 

\subsection{Mixed OU Model}\label{subsec:mixmodel}
\citet{kelly2011} proposed a model for superpositions of OU processes by changing the
scale and magnitude of the process, especially for the X-ray light curves. 
When the number of OU processes is $M$, the model parameters are the lowest and the highest frequencies 
$\omega_L, \omega_H$, noise magnitude $\zeta$ and $\alpha$ that determines the magnitudes of each
process(weight of mixture). It is noted that the model parameters and method of mixing the OU process 
is designed to be able to produce the double-bending PSDs, which can be seen in X-ray variability around 
black holes and also in recent results of QSO variability studies based on the data with more dense cadence such as 
$Kepler$, and ground base telescope data
\citep{mushotzky2011,carini2012,wehrle2013,edelson2014,graham2014,revalski2014,kasliwal2015a,
kasliwal2015b,chen2015,shaya2015,kasliwal2017}. 

When we denote $\omega_k$ as $k$-th value of $M$ with equal ratio division
from $\omega_L$ to $\omega_H$ and the OU process in Equation \ref{eq:ou} with $\tau=\frac{1}{\omega_k}$ 
as $X_k$, the mixture model is written as
\begin{equation}
 X= \sum_{k=1}^M c_k X_k,
\end{equation}
where $c_k$ are magnitudes of $X_k$ that are proportional to
$\omega_k^{1-\frac{\alpha}{2}}$ and that $\sum_{k=1}^M c_k^2 = 1$. This
model is regarded as the summation of various OU process with magnitudes
depending on the frequency.

Then, the covariance function can be expressed in the following way. 
\begin{equation}
\mathcal{R}_{mix OU} = \sum_{k=1}^M \frac{c_k^2 \zeta^2 }{2\omega_k} e^{-\omega_k |\delta t|}.\label{eq:covkelly}
\end{equation}

Since this is a mixture model of OU processes, it can model a wider variety of processes. 
The sum of exponentials in Equation \ref{eq:covkelly} falls more slowly than single exponential function.
This means that the mixed OU process exhibits longer range dependency than a single OU process and is 
better suited for modeling the light curves with long timescale dependencies. 
The power spectral density can also be written in relatively simple form,   
\begin{equation}\label{eq:psdkelly}
\mathcal{P}_{mix OU} = \sum_{k=1}^{M} \frac{c_k^2 \zeta^2}{2\pi} \frac{1}{\omega_k^2 + \omega^2}, 
\end{equation}
where $\omega_k$=$1/\tau_k$ and $\tau_k$ is the damping time scale of $k$-th OU process.  
It should be noted that \citet{kelly2011} assumes that all of the individual OU processes have the same value 
of $\zeta$, for avoiding the degeneracy coming from $c_k^2 \zeta_k^2$, in considering $\zeta_k$ as noise magnitude 
for the $k$-th process.
Also, it does not require much complexity when $M$ is small compared to $N$. 
\citet{kelly2011} succeeded implementing the efficient calculation by obtaining a state-space representation of the light curve 
with about 3,000 data points, and then using the Kalman recursions\citep{brockwell2002}. 

However, it is not easy to determine the appropriate $M$ beforehand. It is unnatural to assume only $M$ processes 
contribute to the variability. Considering $M$ as a hyperparameter and trying to search appropriate $M$ by changing 
$M$ is an option, however it is not realistic because we use the data-set with $10 \sim 10^2$ measurements, 
so it requires too huge amount of time when trying many $M$ values. 
As it requires the time on the order of $O(NM^2)$ to compute the covariance, it is inefficient 
in the case of large $M$ relative to $N$. If we can compute the covariance of mixture OU process model 
with the same time spent for single OU process model, it is very good for applying to large data-set.  
As described in the next coming sections, it is the case for the data-set we are using this time. 

\subsection{Our Model}\label{subsec:ourmodel}
In the previous section, we denoted that it is unnatural to determine the
number of OU processes $M$ beforehand and that computation is not efficient when $M$ is large. 
Thus, we try to set $M\rightarrow \infty$ in
Equation \ref{eq:covkelly}. 

The covariance function of the mixture model described in the Equation \ref{eq:covkelly}
can also be written as 

\begin{equation}\label{eq:ourmodel_cov1}
 \int_{\log \omega_L}^{\log \omega_H}\frac{c(\omega)^2 \zeta^2 }{2\omega} e^{-\omega |\delta t|}d\log\omega.
\end{equation}

if written in the form of integral using $\log(\omega)$. 
As $c(\omega)$ is proportional to $\omega^{1-\frac{\alpha}{2}}$, the equation can be written as 

\begin{equation}\label{eq:ourmodel_cov2}
 \int_{\omega_L}^{\omega_H}\frac{A^2\omega^{2-\alpha} \zeta^2 }{2\omega} e^{-\omega |\delta t|}\frac{d\omega}{\omega}= \int_{\omega_L}^{\omega_H}\frac{A^2\zeta^2}{2}\omega^{-\alpha} e^{-\omega |\delta t|}{d\omega},
\end{equation}

by setting $A$ as a constant of proportionality. 
If we define $x = \omega|\delta t|$, 
\begin{equation}\label{eq:ourmodel_cov3}
\frac{A^2\zeta^2}{2} \int_{\omega_L |\delta t|}^{\omega_H |\delta t|}\left(\frac{x}{|\delta t|}\right)^{-\alpha}
e^{-x}\frac{dx}{|\delta t|}
 =
 \frac{A^2\zeta^2|\delta t| ^{\alpha-1}}{2}\Gamma(1-\alpha,\omega_L|\delta t|,\omega_H|\delta t|).
\end{equation}
and $A$ can be written as 
\begin{equation}\label{eq:ourmodel_cov4}
 1=\int_{\log \omega_L}^{\log \omega_H} A^2
  \omega^{2-\alpha}d\log\omega=\int_{\omega_L}^{\omega_H} A^2 \omega^{1-\alpha}d\omega=A^2\frac{\omega_H^{2-\alpha}-\omega_L^{2-\alpha}}{2-\alpha}.
\end{equation}
because $\Sigma c_{k}^{2} = 1$. 
\par

Additionally the integrand of Equation \ref{eq:ourmodel_cov1} is bounded in the range of 
$\omega_L<\omega<\omega_H,|\delta t|\geq 0$, 
we can exchange the integral and $|\delta t|\to 0$. 
Therefore we can analytically derive the equation by setting $e^{-\omega|\delta t|}=1$ in the integrand as follows, 

\begin{equation}\label{eq:covinf}
\mathcal{R}_{InfOU} = \left\{ \begin{array}{ll}
 \frac{\left(2-\alpha\right)
  \zeta^2}{2\left(\omega_H^{2-\alpha}-\omega_L^{2-\alpha}\right)}|\delta t|^{\alpha-1}\Gamma(1-\alpha,\omega_L|\delta t|,\omega_H|\delta t|) & (|\delta t|>0) \\
\frac{\zeta^2\left(2-\alpha\right)\left(\omega_H^{1-\alpha}-\omega_L^{1-\alpha}\right)}{2\left(1-\alpha\right)\left(\omega_H^{2-\alpha}-\omega_L^{2-\alpha}\right)} & (|\delta t|=0)
                  \end{array} \right.
\end{equation}

as a covariance function, where $\Gamma(1-\alpha,\omega_L|\delta t|,\omega_H|\delta t|)$ is an incomplete
gamma function. 

We can also derive the power spectral density by using the hyper-geometric function. 
The power spectrum of the single OU process with a frequency $\omega$ can be written as 
\begin{equation}\label{eq:ourmodel_psd1}
 \frac{\zeta^2}{2\pi}\frac{1}{\omega^2+x^2},
\end{equation}

Then the power spectrum of our model can be derived by integrating it with $\omega$.  

\begin{equation}\label{eq:ourmodel_psd2}
 \frac{\zeta^2}{2\pi}\int_{\log\omega_L}^{\log\omega_H}\frac{c(\omega)^2}{\omega^2+x^2}d\log\omega= \frac{\zeta^2A^2}{2\pi}\int_{\omega_L}^{\omega_H}\frac{\omega^{1-\alpha}}{\omega^2+x^2}d\omega.\label{eq:psd}
\end{equation}

If we define $\omega=x\tan\theta$, the integration can be written as 
\begin{equation}\label{eq:ourmodel_psd3}
 \frac{\zeta^2A^2}{2\pi}\int_{{\arctan}(\frac{\omega_L}{x})}^{{\arctan}(\frac{\omega_H}{x})}
  \frac{x^{1-\alpha}\tan^{1-\alpha}\theta}{x^2(1+\tan^2\theta)}\frac{xd\theta}{\cos^2\theta}
=\frac{\zeta^2A^2x^{-\alpha}}{2\pi}\int_{\arctan(\frac{\omega_L}{x})}^{\arctan(\frac{\omega_H}{x})}\tan^{1-\alpha}\theta
d\theta,
\end{equation}

As the integrand of equation \ref{eq:ourmodel_psd3} can be written as below, 
\begin{equation}\label{eq:ourmodel_psd4}
 \int \tan^{1-\alpha}(x)dx=\frac{1}{2-\alpha}{\rm Hyper_{2F1}}\left(1,1-\frac{\alpha}{2},2-\frac{\alpha}{2},-\tan^2(x)\right)\tan^{2-\alpha}(x),
\end{equation}
then the spectrum of our model can be described as 
\begin{equation}\label{eq:ourmodel_psd5}
\mathcal{P}_{infOU} = \frac{\zeta^2A^2x^{-\alpha}}{2\pi(2-\alpha)}\left({\rm Hyper_{2F1}}\left(1,1-\frac{\alpha}{2},2-\frac{\alpha}{2},-\left(\frac{\omega_H}{x}\right)^2\right)\left(\frac{\omega_H}{x}\right)^{2-\alpha}-{\rm Hyper_{2F1}}\left(1,1-\frac{\alpha}{2},2-\frac{\alpha}{2},-\left(\frac{\omega_L}{x}\right)^2\right)\left(\frac{\omega_L}{x}\right)^{2-\alpha}\right),
\end{equation}
The hyper-geometric function ${\rm Hyper_{2F1}}\left(a,b,c,z\right)$ is defined as below,  
\begin{equation}\label{eq:ourmodel_psd6}
{\rm Hyper_{2F1}}\left(a,b,c,z\right)=\sum_{n=1}^\infty \frac{(a)_n(b)_n}{(c)_n} \frac{z^n}{n!}
\end{equation}
where $(q)_n$ is the Pochhammer symbol, which is defined by 
\begin{equation}\label{eq:ourmodel_psd7}
(q)_n = \left\{ \begin{array}{ll}
         1 & (n=0) \\
         q(q+1)...(q+n-1) & (n>0)
        \end{array} \right.
\end{equation}

This model is regarded as the infinite superpositions of OU process with scales from 
$\omega_L$ to $\omega_H$. Thus, it is more natural than \citet{kelly2011}'s model which assumes  
that only finite discrete OU processes to contribute the variability. Since we can omit $M$ in the
covariance function as shown in Equation \ref{eq:covinf}, 
it does not require $M$ times computation in the covariance calculation, which is leading the faster calculation.

\subsection{Practical Implementation and Model Calculation}\label{subsec:ourmodel_imple_calc}
We infer the model parameters by the following procedures. First, we estimate covariances for 
all measurement sets, by using each measurements and errors for sampling the data, to produce 
the covariance matrix, as described in \ref{subsec:ourmodel}. 
After calculating the likelihood by the covariance, we try to get posterior distribution of 
model parameters using MCMC sampler, and infer the best parameter values by getting medians of 
each parameter distributions. 
We use similar prior distribution to the prior used for the analysis by \citet{kelly2011}, but there is 
a difference in the range of $\alpha$, which is $-2 < \alpha < 0$ in \citet{kelly2011}'s work. 
In our calculation for {\rm{MAP}} estimation, for the prior distribution we assume the uniform distribution 
for $\alpha$ in the range of $-3 < \alpha < 3$.
It is because the likelihood values are systematically larger than the case of \citet{kelly2011}'s 
definition, for the objects which seem to have different PSDs inferred by DRW model.   
For $\zeta$ we use uniform prior in the range of $\zeta > 0$, and also assume the uniform prior 
on $\omega_L$ and $\omega_H$, which satisfy the following relations, 
$\omega_{min} \ll \omega_L \ll \omega_{max}$, and $\omega_L \ll \omega_H \ll \omega_{max}$. 
The upper and lower limits on the characteristic time scales, 
$\tau_{max}=1/\omega_{min}$ and $\tau_{min}=1/\omega_{max}$ are chosen to be 10$^5$ and 10$^{-2}$ days, respectively. 
We set the range of $\omega$ is about one order larger and smaller than the range of time intervals of 
the measurements. It is for avoiding the effects coming from boundary conditions in time for our analysis, 
and we only must see the parameter ranges which can be covered by the observations, if we want to interpret 
the resultant PSDs. 

We use a part of {\rm{JAVELIN}} package
\footnote{http://www.astronomy.ohio-state.edu/~yingzu/codes.html\#javelin}
\citep{zu2011,zu2013} for producing the light curve of the object with 
the given data points (usually 40 to 120 in our analysis), which is based on MCMC processing. 
In this package MCMC library $emcee$
\footnote{http://dan.iel.fm/emcee/current/}\citep{foreman-mackey2012} is used for random sampling. 

As the covariance function is an incomplete gamma function, we use the GNU Science Library 
for the faster calculation. As the {\rm{JAVELIN}} package is 
written in $python$ language, we use $python$ as the language for our software development.  

As the calculation for each object can be processed in parallel, we implemented the multi-process(parallel)   
calculation in our model analysis. Our model calculation takes about two to three minutes per object in 
single thread(depends on the number of data points), and we complete our model calculation in about five days 
for all 67,507 objects in our sample with 20 threads processing with CPU of Intel Xeon E5507(2.27GHz).

\section{Data} \label{sec:data}
\subsection{SDSS Stripe 82 Photometric Catalog}\label{subsec:photocat}
We use the $gri$ photometric information from the \citet{ivezic2007} variable 
source catalog, downloaded from the site
\footnote{http://www.astro.washington.edu/users/ivezic/sdss/catalogs/S82variables.html}. 
The catalog contains light curve information of 67,507 variable source 
candidates identified in Sloan Digital Sky Survey (SDSS) Stripe 82 area, based on SDSS Data Release 7
(DR7)\citep{abazajian2009}. 
Their criteria for selecting the candidates are 
\begin{enumerate}
\item{unresolved source in imaging data, at least one band with photometric error below 0.05 mag}
\item{processing flags BRIGHT, SATUR, BLENDED, or EDGE are not set, meaning the measurement performed successfully}
\item{at least 10 observations in the $g$ and $r$ bands}
\item{the median $g$ band magnitude brighter than 20.5}
\item{root-mean-square scatter $>$ 0.05 magnitude and $\chi^{2}$ per degree 
of freedom larger than three in both $g$ and $r$ bands, which means the object is 
'statistically' variable in brightness (see the discussions in \citet{sesar2007} for the details)}
\end{enumerate}
This catalog is more extensive than the catalog of \citet{sesar2007}, 
because it includes both SDSS-I and SDSS-II, while the \citet{sesar2007}'s catalog 
was based only on SDSS-I and therefore we decided to use this catalog for our research. 
Though this catalog also contains $u$ and $z$ band data, we did not use it for 
our study because of their lower signal-to-noise ratio(S/N) than those three ($gri$) bands data.   
The number of measurements in $r$ band is distributed in the range between $\sim$10 to 140, 
with maximum peak at around 60. 

\subsection{SDSS DR12 Spectroscopic Catalog}\label{subsec:speccat}
We used SDSS Data Release 12(DR12) \citep{alam2015} data for identifying the spectroscopically 
confirmed variable source candidates in the photometric catalog mentioned in \ref{subsec:photocat}. 
On SDSS CasJobs system for DR12 data\footnote{https://skyserver.sdss.org/CasJobs/}, 
we used 'dbo.SpecPhoto' table to select the spectroscopically confirmed objects in the area of Stripe 82. 
We specified the area by $|\delta|<$1.266 deg. and RA in the range 20$^{h}$34$^{m}$ to 4$^{h}$00$^{m}$. 
The number of objects extracted from the table is 221,656, including 25,674 QSOs, 78,761 stars and 
117,221 galaxies. 

Spatial matching with the photometric catalog, based on $r$ band data, is performed for the 
coordinates(photora,photodec) in the table, by the criterion of the difference of the coordinates 
between photometric and spectroscopic catalogs within 1 arcsec. We identified 11,908 matched objects 
and they are 8,105 QSOs, 3,379 stars and 424 galaxies, based on the spectroscopic classification 
on the table 'dbo.SpecPhoto', described as 'QSO', 'STAR' and 'GALAXY' in column 'Class', respectively.  
There are 17 sources with multiple sources in the searched area, 
and we selected the nearest one as the counterpart in the cases. 

It is noted that most spectroscopically confirmed objects in the photometric catalog 
have measurements over 40. 
For the validation of selection ability of QSOs from stars based only on their variability, 
we need a sample with enough number of photometric measurements and spectroscopic identifications.  
Thus we decide to use the objects which are spectroscopically confirmed and have measurements number 
over 40 with good accuracy, with the error of each measurement less than 0.05 magnitude, 
for our test of QSO selection, which hereafter is called a 'good sample' hereafter, 
although we calculate the model parameters on all objects to investigate the effects of measurement 
numbers on our analysis. 
The total number of objects in the 'good sample' is 9,855, with 7,714 QSOs and 2,141 stars. 

\begin{figure}[htb!]
\figurenum{1}
\gridline{\fig{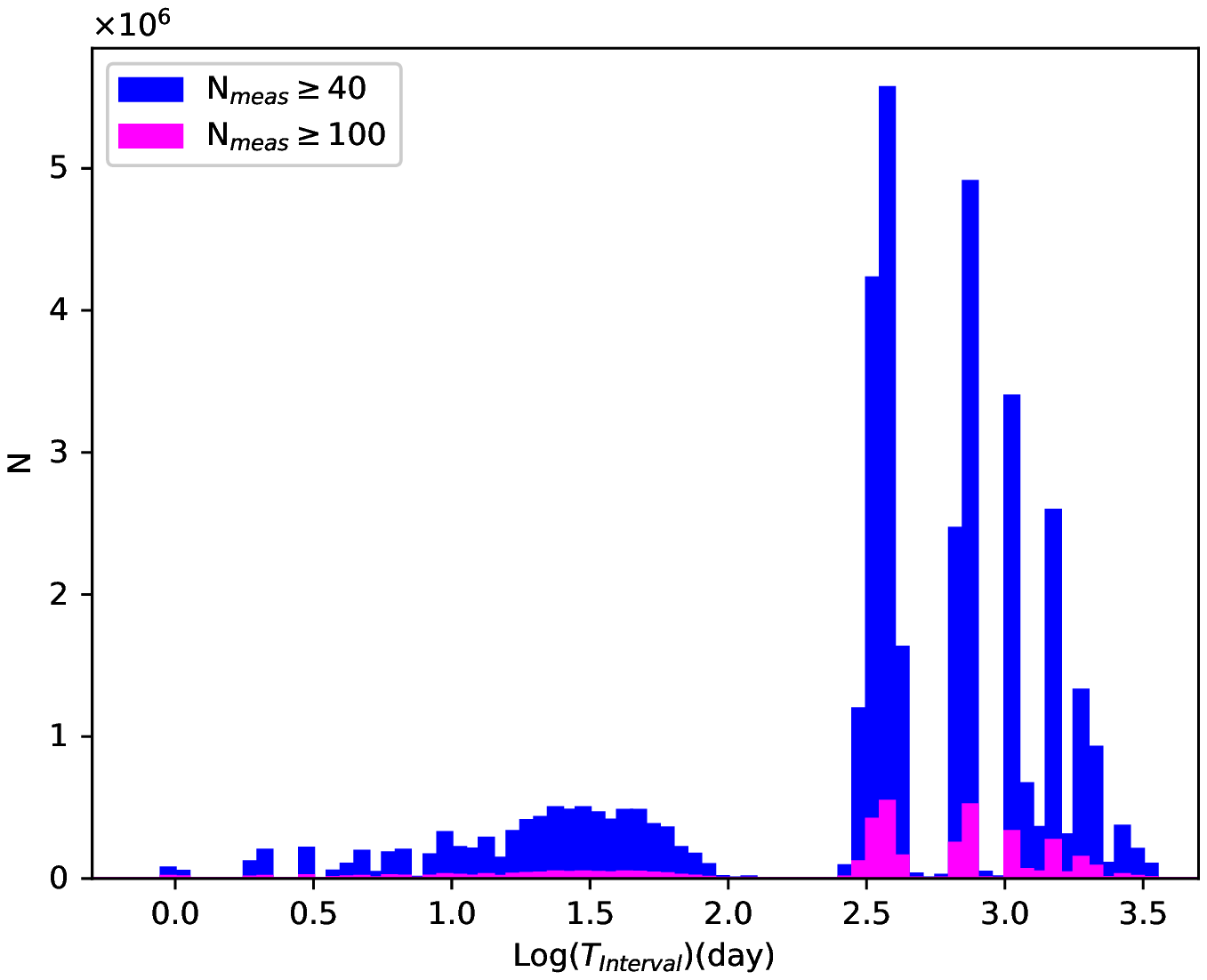}{0.45\textwidth}{(a)}
          \fig{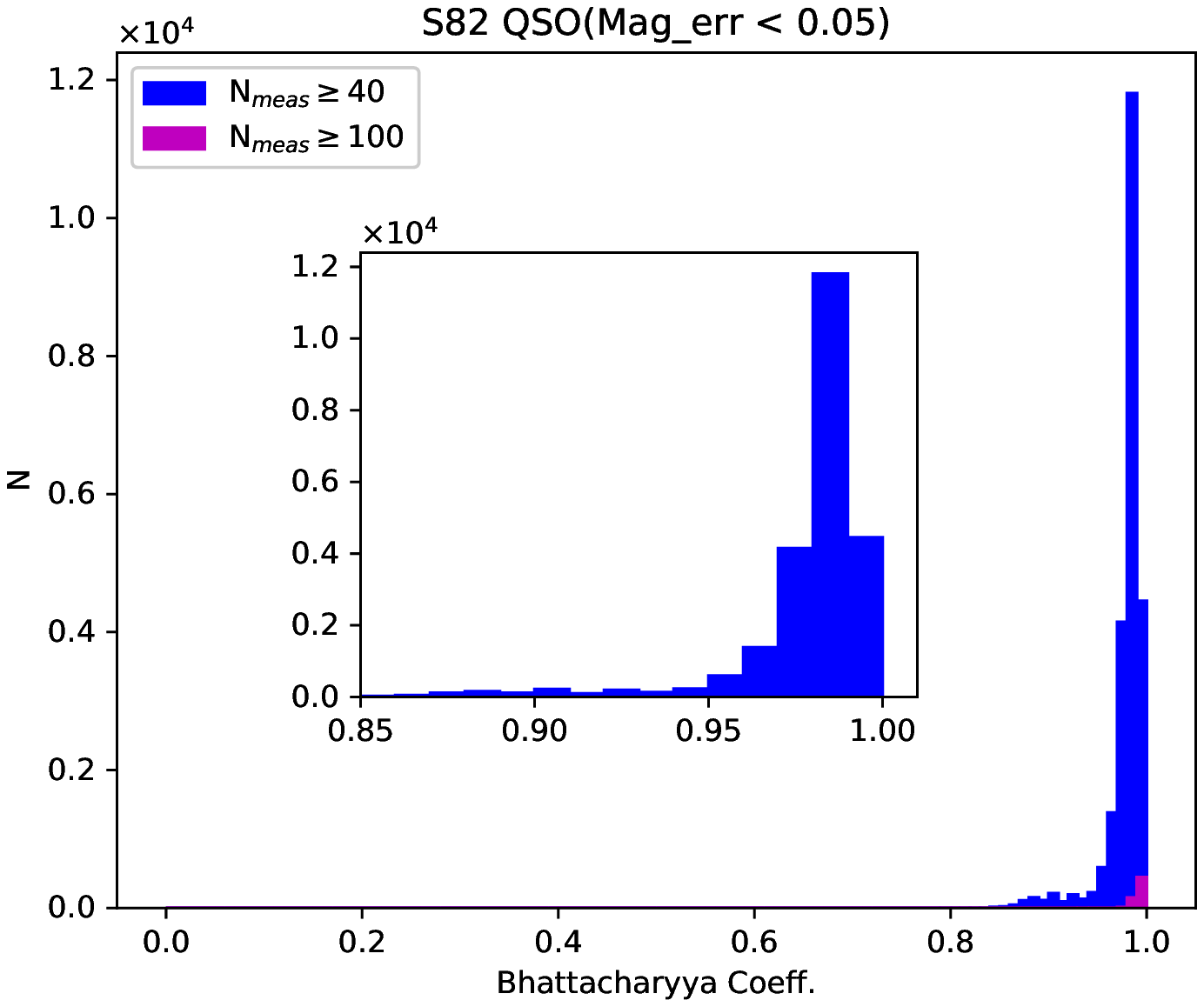}{0.45\textwidth}{(b)}}
\caption{(a)The histogram which shows the 'stacked' cadence of photometric data of 
the 'good sample'. The time intervals are in log-scale. Blue histogram shows for the 
objects with measurements $\geq$40, and magenta for those with $\geq$100 
measurements. It is clear that the cadence is not heavily dependent on the number 
of measurements.    
(b)The histogram of $BC$s for all objects in the 'good sample'. The zoom-up around 
$0.85 < BC < 1.0$ is shown in the inner panel.   
\label{fig:meas_num}}
\end{figure}

One important features that affects the time series analysis using irregularly and sparsely 
sampled data is the cadence of each object and its similarity. 
We consider that the cadences in the sample are similar to each other, 
if the distribution of the sets of time intervals for each object, which is the histogram 
of time intervals, are similar for whole sample.  
To check the cadence of the 'good sample', we calculated time intervals of multiple measurements 
for each object in the sample, and produced the histogram as shown in Figure \ref{fig:meas_num}(a).  
We used 100 bins equally divided between -1 to 4 on log-scaled time intervals in days, 
and stacked all the time intervals of each object to produce a stacked cadence template histogram. 
It is clear that the minimum time intervals are less than $\sim$1 day, and $\sim$10 years at the maximum, 
and we can obtain a certain level of information on variability about such a time-scale range. 

For investigating the similarity of the cadences of our sample, we used the Bhattacharyya 
coefficient(hereafter BC)\citep{bhattacharyya1943}, which is frequently used for checking 
the similarity of two histograms. 
The definition of BC can be written as  
\begin{equation}
BC(p,q) = \sum_{i=1}^n \sqrt{p_i q_i}.\label{eq:bhattacharyya}
\end{equation}
where $p_i$ and $q_i$ are normalized count in $i$-th bin of two histograms. 
In this case they are histograms for stacked cadence template and for single object, respectively. 
If both histograms are the same, $BC$ value will be one, and decrease as dissimilarity increases. 

The distribution of $BC$s for the 'good sample', shown in Figure \ref{fig:meas_num}(b), 
means that the cadences are very similar, as the most of all objects show the values more than 
0.95. This feature of the data set is suitable for detecting 
different types of intrinsic variability as we can decrease the effect coming from irregular 
and sparse sampling. 
It should be noted that the results are the same for histogram with 1 day bin size, which means 
that the cadence greater than 1 day long are very similar for whole sample.  

The similarity of the cadences are easily imagined because SDSS Stripe 82 data are taken through the 
time-delay-and-integrate (TDI) scanning mode\citep{gunn1998}, which enable homegeneous scanning of the sky 
in each observation night. 

As is already mentioned, we cannot investigate the details of intrinsic variability since the data 
suffer from irregular and sparse time sampling. Considering the similarity of the cadence, however, 
the data provide enough information for detecting the relative differences of the photometric variations, 
and they can be powerful to separate QSOs from stars. 
It is meaning that the effects coming from the irregular and sparse time sampling 
will emerge in the same way to the results for all sample objects 
and intrinsic features can be distinguished relatively. 

\section{Results} \label{sec:results}
\subsection{Fitted Light Curves and Derived Parameters} \label{subsec:fitlc_model}
By applying our model to the 67,507 variable sources, we derived four parameters 
($\omega_H$, $\omega_L$, $\zeta$, $\alpha$) for all sources. 
These parameters are estimated by taking the median of posterior distributions, and lower 
and higher errors are estimated by getting the difference between the median and 16\% or 84\% 
quantile values, respectively. 
We calculated them on $g$, $r$, and $i$ band data independently. As the ability of distinguishment of QSOs 
from stars are very similar among the data-set on three filters, we limit our description only on $r$ band 
data in the following. 

\begin{figure}[htb!]
\figurenum{2}
\gridline{\fig{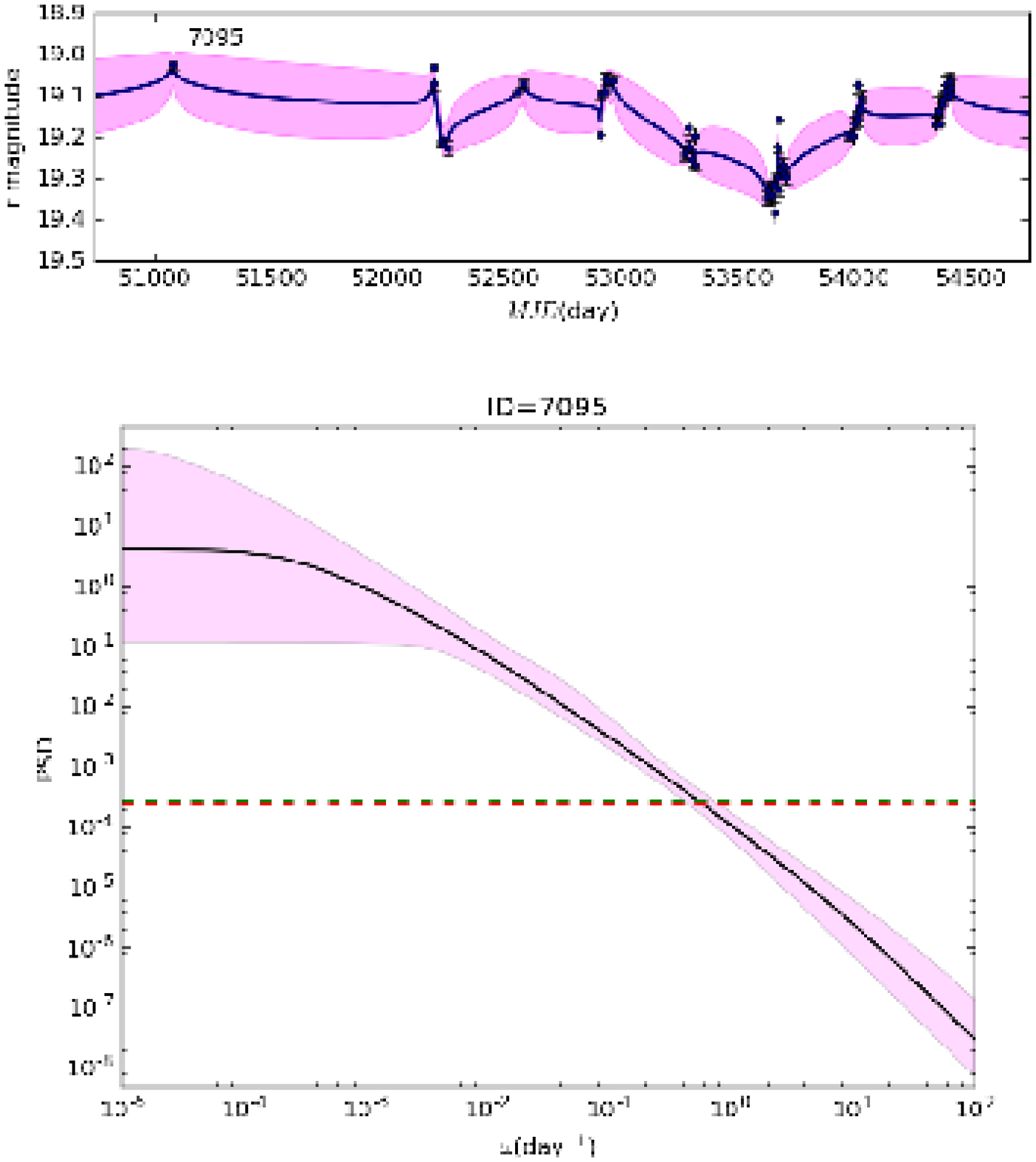}{0.45\textwidth}{(a)}
          \fig{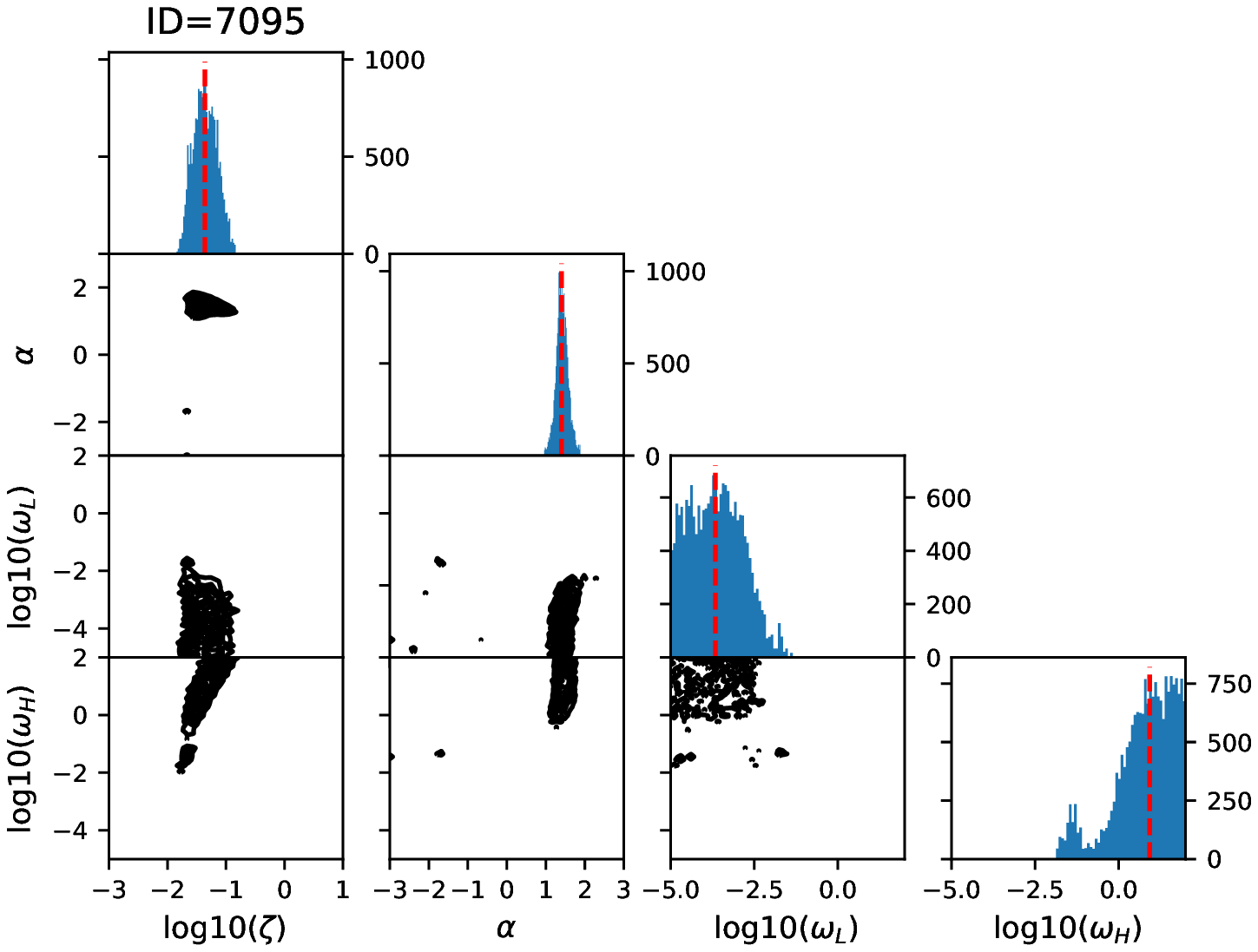}{0.55\textwidth}{(b)}}
\caption{One result of our model calculation for a QSO(ID number 7095). 
(a)Upper left: It shows the light curve with shaded area which presents the 68\% confidence range  
for the inferred light curve. The results of photometric measurements are shown by filled circle 
with error bars. 
Lower left: This shows the inferred PSD with 90\% confidence range with shaded area. 
The horizontal dashed lines show the median(red) and average(green) values of errors 
on photometric measurements. The vertical blue dashed lines show inferred 
$\omega_L$ and $\omega_H$, respectively.  
(b)Pair plot which shows the posterior distributions of our model parameters, 
log $\zeta$, $\alpha$, log $\omega_L$, and log $\omega_H$ from left to right, respectively. 
The histograms show the posterior distributions of each parameters, with median values 
indicated by the red dashed lines. 
Contour maps showing the distributions of the posteriors on each set of our model parameters 
($\zeta$, $\alpha$, $\omega_L$, $\omega_H$) about the object in MCMC chains.
\label{fig:infmodel_analysis_qso}}
\end{figure}

\begin{figure}[htb!]
\figurenum{3}
\gridline{\fig{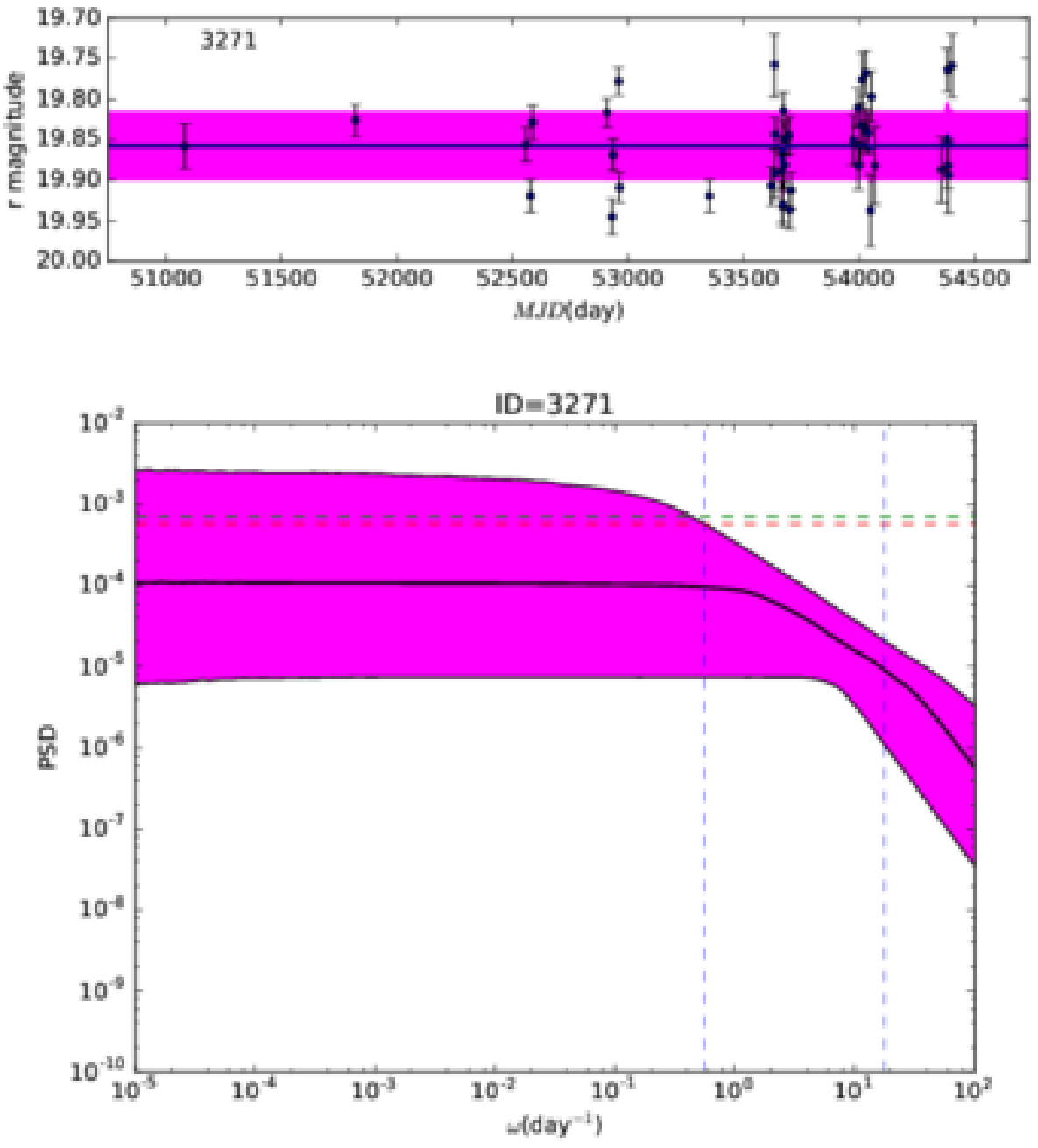}{0.45\textwidth}{(a)}
          \fig{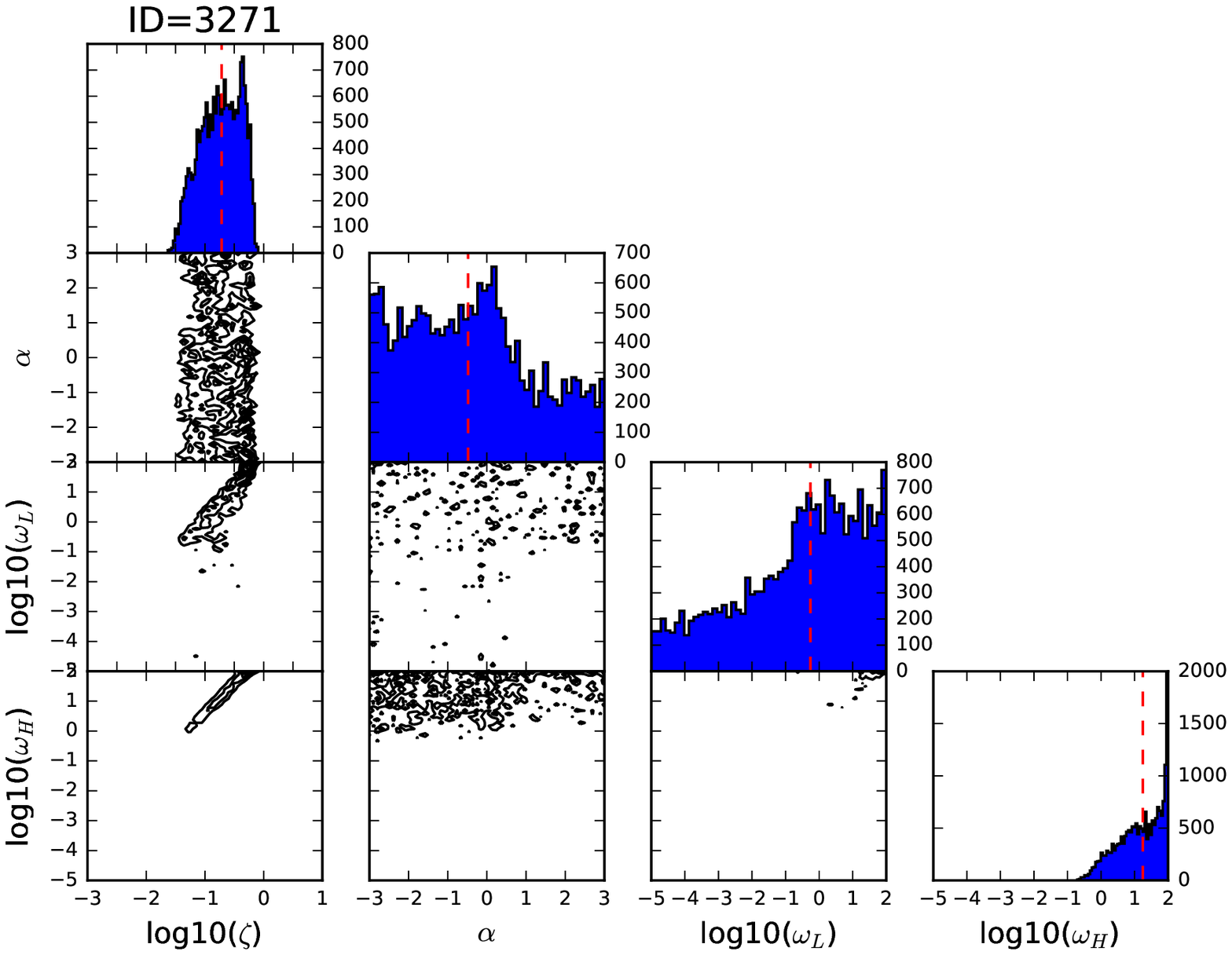}{0.55\textwidth}{(b)}}
\caption{The same one of Figure \ref{fig:infmodel_analysis_qso} for a star(ID number 3271).
\label{fig:infmodel_analysis_star}}
\end{figure}

In Figures \ref{fig:infmodel_analysis_qso} and \ref{fig:infmodel_analysis_star}, 
we show the inferred light curves for one of the spectroscopically confirmed QSOs and stars, respectively. 
These figures also show their posterior distributions of inferred parameters in histograms, 
on the data sets of posteriors in contours, and their inferred power spectral densities (PSDs), 
which will be described on the details in \ref{subsec:psd}. 
The shaded part of the light curves show the range of 68\% confidence level of our calculation 
based on MCMC, and the solid lines mean the inferred light curves, derived from median of the 
20,000 chain results.  

On the most spectroscopically confirmed QSO, we can get well fitted inferred light curves, 
which are characterized by variation with long time scale($e.g.$ small frequency) 
(Figure \ref{fig:infmodel_analysis_qso}(a)). They are characterized by small $\omega_L$ and 
$\omega_H$. 

On the other hand, we cannot have a good fit on stars light curve, with flat light curves 
without tracing the measurements, as shown in Figure \ref{fig:infmodel_analysis_star}(a). 
It is however not the failure of our model fitting because our model uses only a type of PSD 
assuming the QSO/AGN like variability. In fact the inferred PSD has larger $\zeta$ compared to 
QSO/AGN, and flat shape in small $\omega$ range which means the variability is like white noise. 
As the most variable stars are periodically variating their brightness, it is consistent to the   
the fact. Most importantly inferred parameters are good for the purpose of separating stars 
from QSOs in the parameters space, as shown in next coming sections.  

There are also 37 spectroscopically confirmed QSO in the 'good sample' without successful light curve fits.  
They cannot be distinguished from most stars in optical variability using our model. 
On the other hand, we identify 199 QSO-like light curves on the objects classified as star on the 
SDSS spectra. We investigated carefully their optical spectra by visual inspection and determined 16 of them 
are possibly misidentified as 'STAR'. Many misidentified objects have relatively weak emission 
lines in their spectra and are possibly categorized as weak emission line QSOs\citep{plotkin2010}.

We should note that the inferred parameter $\omega_L$ for many QSOs are less than $\sim$1/4000 days$^{-1}$, 
the maximum duration limit of the data($\sim$10 years). It is because we set the prior of $\omega_L$ 
as uniform distribution in the range of $-5 < {\rm{log}}\omega_L < 2$. 
The values are thus the results of fitting our model to the data with time-scale shorter than $\sim$4,000 days, 
and $\omega_L$ smaller than 4000$^{-1}$ day$^{-1}$ is only the extrapolation based on our model. 
We should emphasize here that the values of $\omega_L$ are meaningful, as our purpose is only to identify QSOs, 
not to infer PSDs correctly, by using a model with approximate PSD expression for describing variability of QSOs. 
We also mention briefly the interpretation of the inferred PSDs in \ref{subsec:agn_psd_param}. 

For confirming the consistency of our model with \citet{kelly2011}'s finite OU mixture model, 
we checked the value of information criteria on the results. 
We use BIC(Bayesian Information Criterion)\citep{schwarz1978} for this check. 
It is defined as the following equation, 

\begin{equation}\label{eq:bic}
BIC = k\ln(n) - 2\ln(L)
\end{equation}

where $L$ is the maximum value of likelihood function, $n$ is the number of measurements, 
and $k$ is the number of estimated parameters in the model.

Information criteria are a common and useful mechanism for ranking a set of models.  
In time series analysis BIC and/or AIC(Akaike Information Criterion)\citep{akaike1973} are used so frequently, 
which is based on the maximum-likelihood estimate of the parameters. 
The BIC/AIC provide estimates of the relative information lost in using a model to represent the underlying process 
that generated the data. As we used Bayesian inference for our analysis, we use BIC for the check. 

\begin{figure}[htb!]
\figurenum{4} 
\gridline{\fig{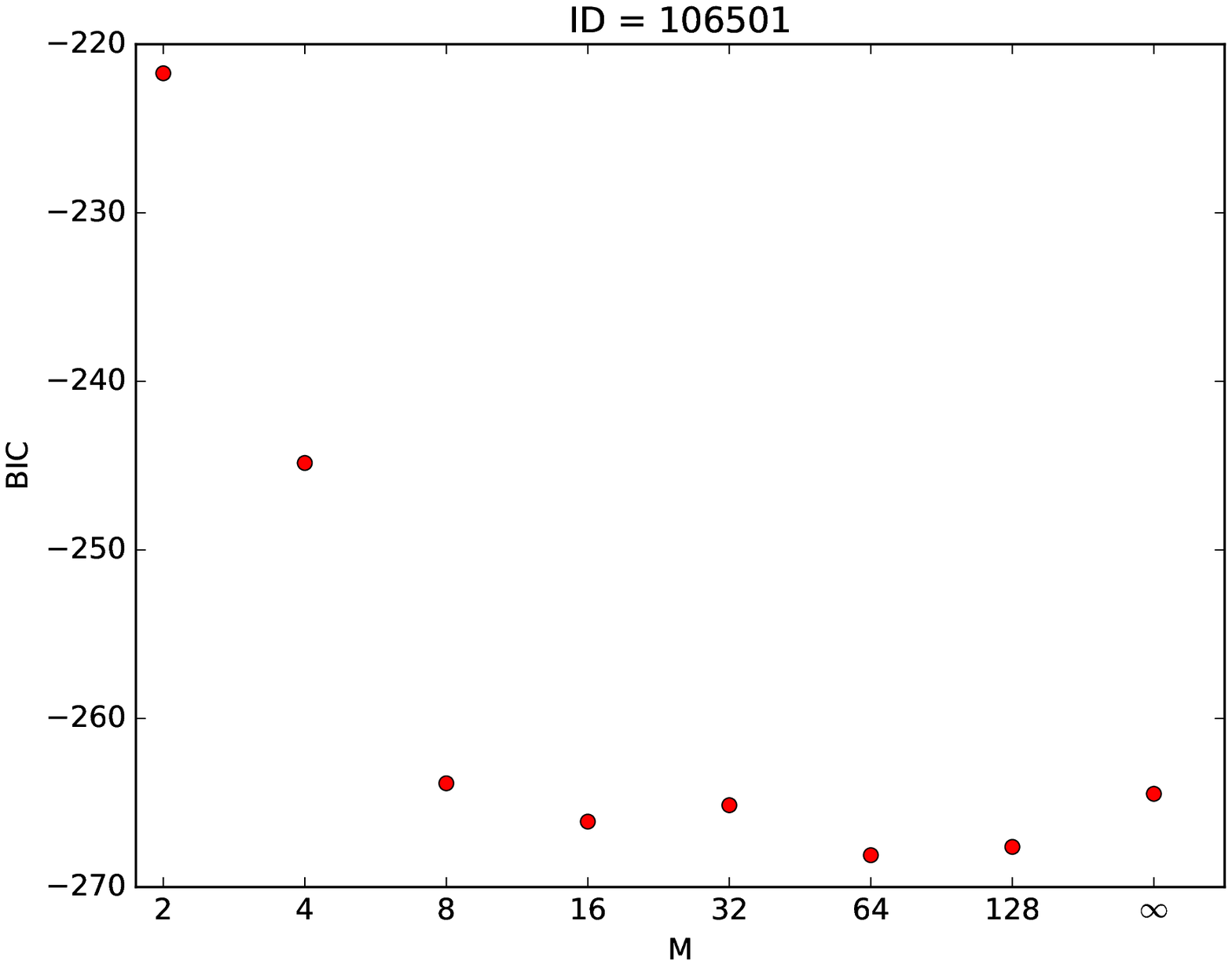}{0.3\textwidth}{(a)}
          \fig{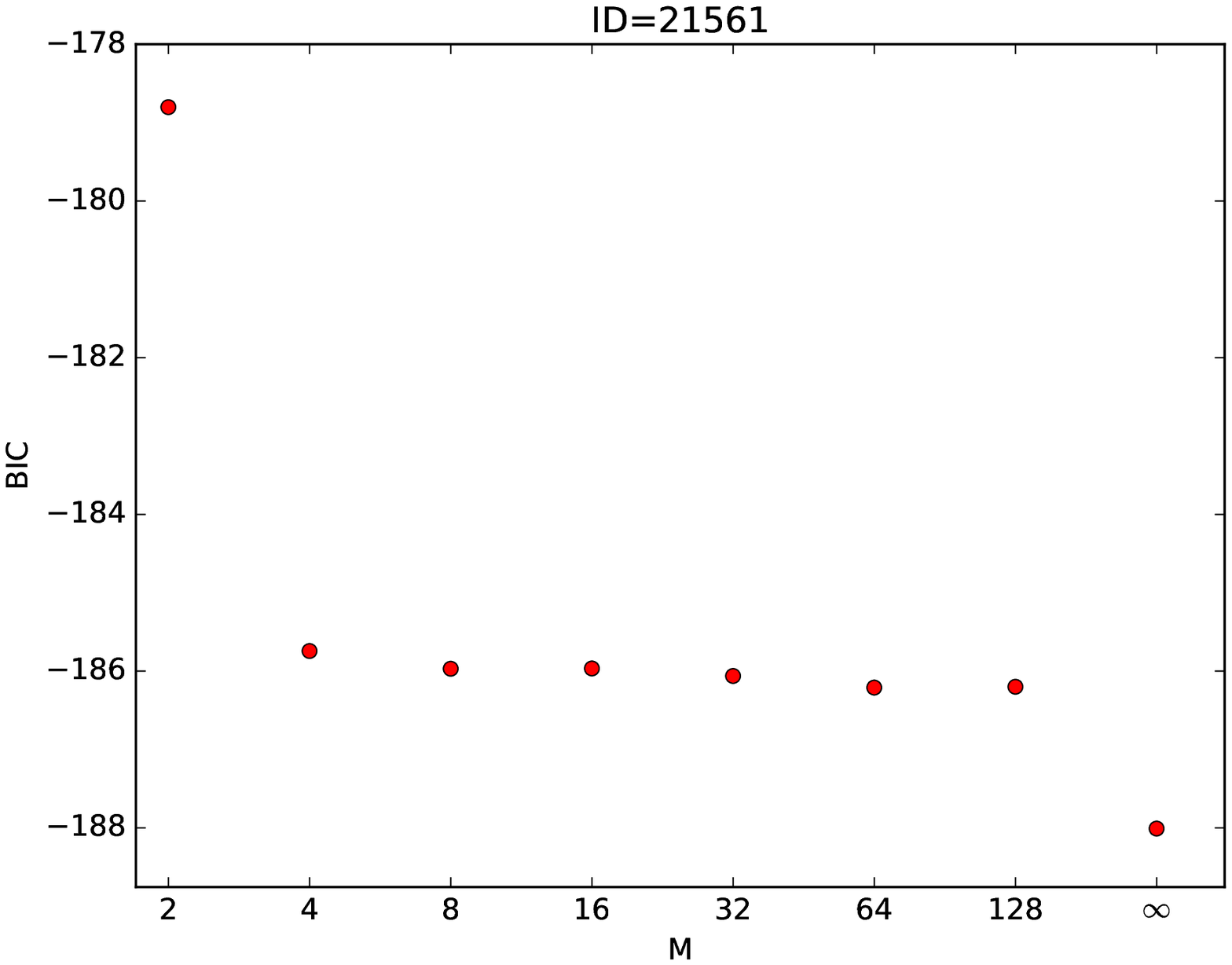}{0.3\textwidth}{(b)}
          \fig{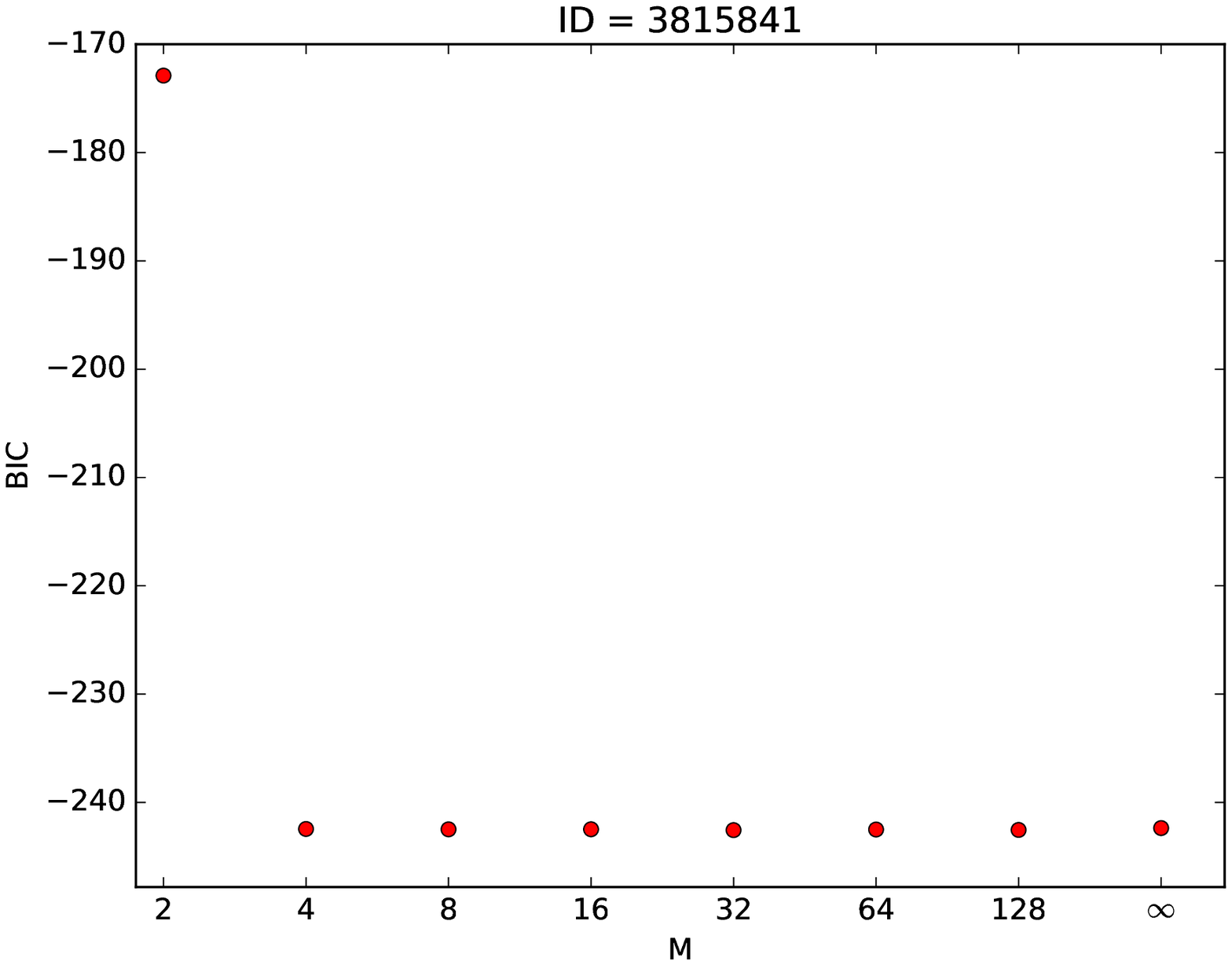}{0.3\textwidth}{(c)}}
\caption{BICs dependencies on the number of mixed OU processes about 3 sample objects. 
BIC on $M\rightarrow \infty$ show the comparable values to those for $M>$30.  
 \label{fig:bics_qsos}}
\end{figure}

In Figure \ref{fig:bics_qsos}, we show 3 samples of BICs distributions on $M$ about 
spectroscopically confirmed QSOs. There are various tendencies on the distribution, however 
we can see our infinite OU mixture model provides the same level of the goodness of the fit 
compared to \citet{kelly2011}'s finite OU mixture model with large mixture numbers. 
It is also consistent with the suggestion by \citet{kelly2011} that the number of mixed OU 
processes should be larger than 30 to describe observed AGN's light curves sufficiently. 
We should note here that the numbers of mixed OU processes($M$) which provide the best BIC on mixed OU model 
for the spectroscipically confirmed QSOs in 'good sample' change object to object, 
although most of them are $M=$32,64,128 or $\infty$. 

\begin{figure}[htb!]
\figurenum{5} 
\gridline{\fig{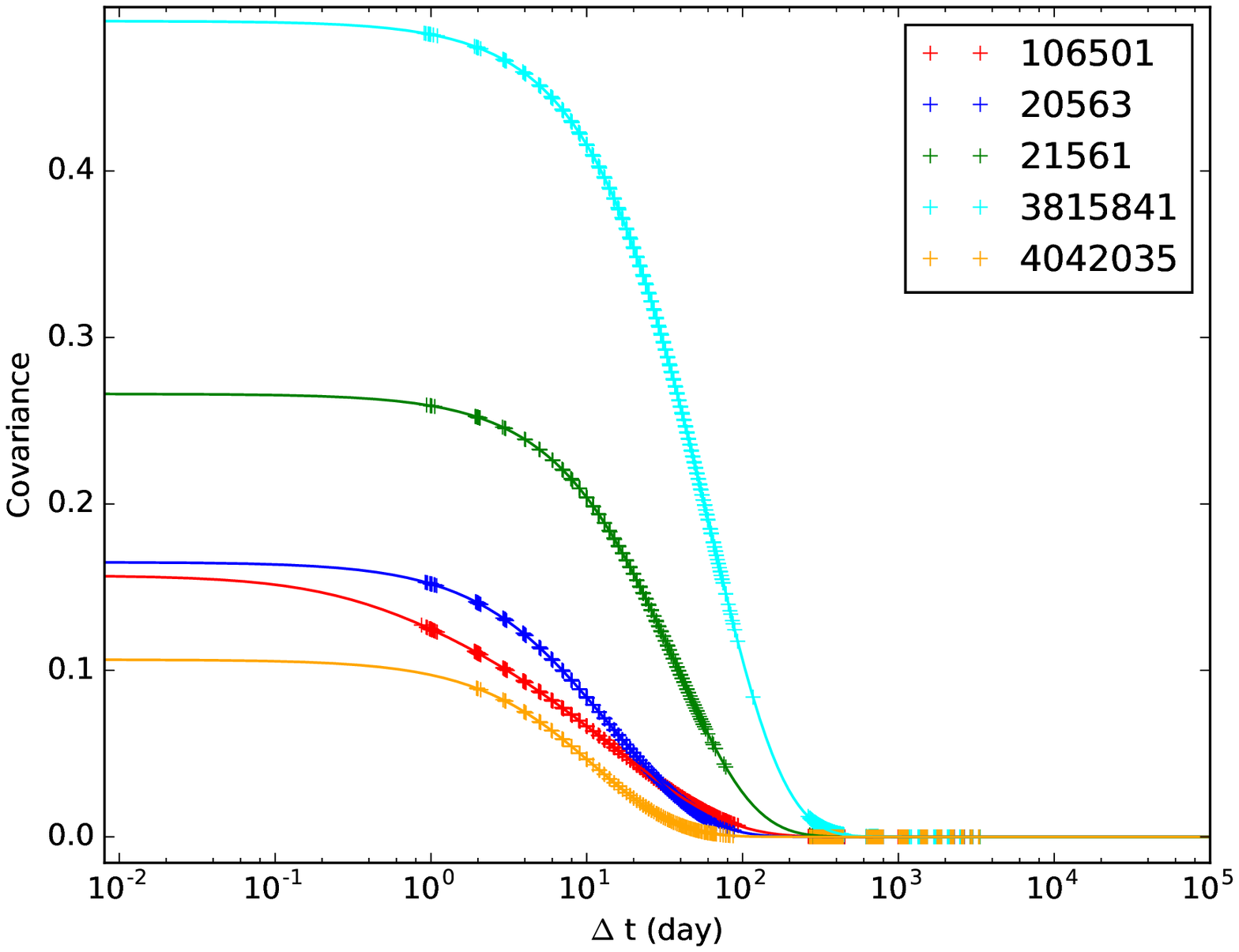}{0.45\textwidth}{(a)}
          \fig{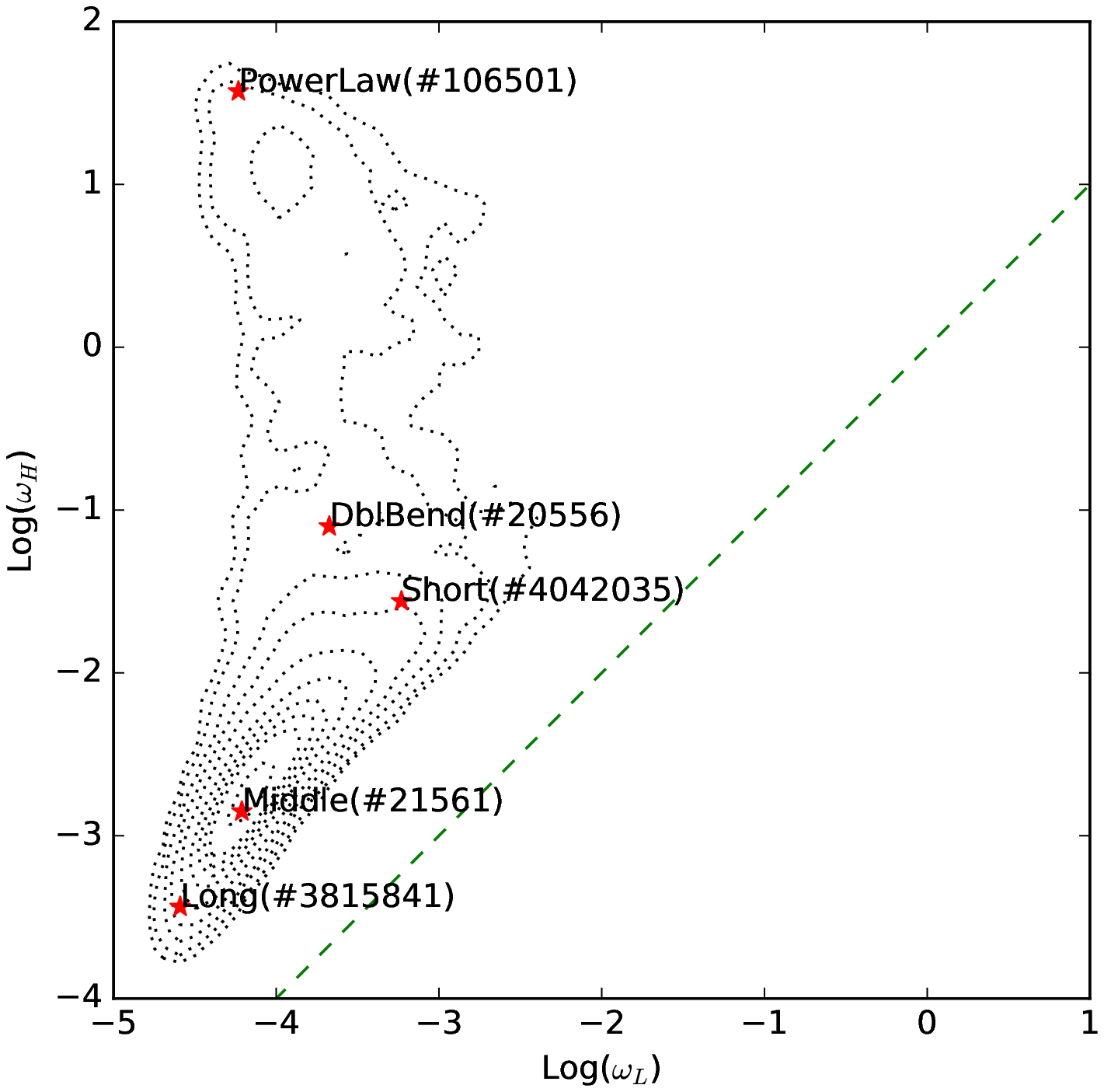}{0.45\textwidth}{(b)}}
\caption{(a)Covariance functions for 5 sample objects with different types of variabilities.
(b)The contours with dotted lines show the distribution of spectroscopically confirmed QSOs 
in our 'good sample' on $\omega_L$ - $\omega_H$ plane. 
The types of variabilities are shown with ID number of the representative objects. 
 \label{fig:covariance_function}}
\end{figure}

For checking the flexibility of our model to AGN/QSO variability, we show the covariance functions 
in Figure \ref{fig:covariance_function}(a). 
The crosses are covariance values calculated by the observation data set. 
We select 5 objects which are representative of our model parameter space 
as shown in Figure \ref{fig:covariance_function}(b). 
We named them as 'Long', 'Middle', 'Short', 'DblBend' and 'PowerLaw'.

'Long', 'Middle' and 'Short' represent the difference of variability time scales, 
and they show smaller difference between $\omega_{H}$ and $\omega_{L}$. 
'DblBend' and 'PowerLaw' represents the shapes of PSD with double-bending and power-law which have 
much difference between $\omega_{H}$ and $\omega_{L}$. 
The difference between 'DblBend' and 'PowerLaw' is whether $\omega_H$ is lower or higher than 1 day$^{-1}$, 
which corresponds to the minimum of time intervals of the data.   

It is clear that the covariance functions can describe well the difference of 
variabilities by changing their height and shape, especially in short time scale 
range.  

\subsection{PSD Calculation and Derived Parameters} \label{subsec:psd}
\begin{figure}[htb!]
\figurenum{6}
\gridline{\fig{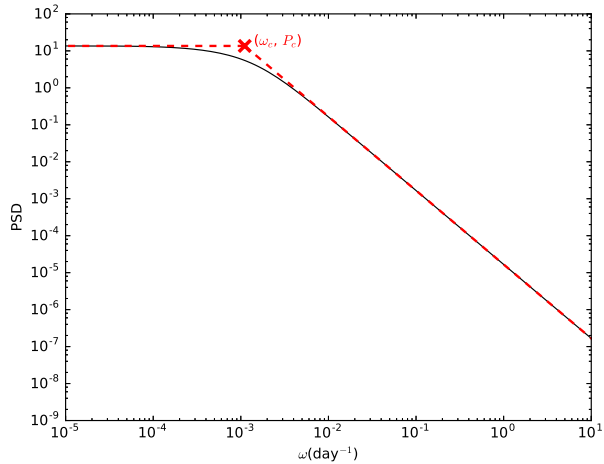}{0.5\textwidth}{}}
\caption{The schematic figure showing how the values of PSD's parameters $\omega_{c}$ and $P_{c}$ 
are derived. \label{fig:infmodel_psd3dparam}}
\end{figure}

For investigating the details of the variability, we derive the power spectral density (PSD) 
for each object. The method to calculate the PSD is described in \ref{subsec:ourmodel},
and PSDs for a sample QSO/AGN and star are shown in Figure \ref{fig:infmodel_analysis_qso} and 
Figure \ref{fig:infmodel_analysis_star}. 
We can see some typical features in PSDs for QSOs and stars. For example, many QSOs show the 
PSDs with smoothly changing slopes, the turning points appears at the border of 
white noise (slope $\beta$=0) and red noise($\beta$=$-$2), when we express the PSD $ \propto \omega^{\beta}$.    
On the other hand, stars' PSD inferred by our model has a bending point at larger $\omega$ range
(log $\omega > 0$), with flat spectra in smaller $\omega$ range, which is consistent to white noise.  
Since our model originally needs four parameters for describing PSD for the variability of QSOs,
it is not easy to visualize the behavior of inferred PSDs among different celestial objects, 
especially QSOs and stars. We therefore introduce the parameters, which show the typical frequency 
and height of PSDs. 
As the PSD derived from our model has the features that the slope will be asymptotic to 
0 in $\omega \to -\infty$, and to -2 in $\omega \to \infty$, we define two asymptotic lines for 
the PSD and derive the ($\omega$, $P(\omega)$) at the crossing point of these lines, as ($\omega_{c}$, $P_{c}$), 
and then we try to describe our model with three parameters $\omega_{c}$, $P_{c}$ and $\alpha$
(Figure \ref{fig:infmodel_psd3dparam}). 
It means that $\omega_{c}$, $P_{c}$ and $\alpha$ are expressing the features of PSDs, which are corresponding to 
the absolute position of typical time scale, the amplitude, and difference of maxmum and minimum time scales.    
\par
We define the feature point on the power spectrum derived from our model as the crossing point of the asymptotes 
in the cases of $\omega \rightarrow 0$ and $\omega \rightarrow \infty$. 
When $\omega \rightarrow 0$, the power spectrum can be written as below.
\begin{equation}\label{eq:ourmodel_fpt1}
 \frac{\zeta^2A^2}{2\pi}\int_{\omega_L}^{\omega_H}\frac{\omega^{1-\alpha}}{\omega^2}d\omega
=\frac{\zeta^2A^2}{2\pi}\int_{\omega_L}^{\omega_H}\omega^{-1-\alpha}d\omega
=\frac{\zeta^2A^2}{2\pi}\frac{\omega_H^{-\alpha}-\omega_L^{-\alpha}}{-\alpha},
\end{equation}
On the other hand in case of $\omega \rightarrow \infty$, we can derive the spectrum as below. 
\begin{equation}\label{eq:ourmodel_fpt2}
 \frac{\zeta^2A^2}{2\pi}\int_{\omega_L}^{\omega_H}\frac{\omega^{1-\alpha}}{x^2}d\omega
=\frac{\zeta^2A^2}{2\pi}\frac{1}{x^2}\int_{\omega_L}^{\omega_H}\omega^{1-\alpha}d\omega
=\frac{\zeta^2A^2}{2\pi}\frac{1}{x^2}\frac{\omega_H^{2-\alpha}-\omega_L^{2-\alpha}}{2-\alpha},
\end{equation}
By satisfying the both formulas simultaneously, the coordinate of crossing point is derived as 
\begin{eqnarray}
x &=&
 \sqrt{\frac{-\alpha}{\omega_H^{-\alpha}-\omega_L^{-\alpha}}\frac{\omega_H^{2-\alpha}-\omega_L^{2-\alpha}}{2-\alpha}} \equiv \omega_{c} \label{eq:ourmodel_fpt3}\\
y &=& 
 \frac{\zeta^2}{2\pi}\frac{2-\alpha}{\omega_H^{2-\alpha}-\omega_L^{2-\alpha}}\frac{\omega_H^{-\alpha}-\omega_L^{-\alpha}}{-\alpha} \equiv P_{c}\label{eq:ourmodel_fpt4},
\end{eqnarray}
, and we can define the point as the feature point. 

\begin{figure}[htb!]
\figurenum{7}
%\gridline{\fig{fig7.eps}{0.9\textwidth}{}}
\gridline{\fig{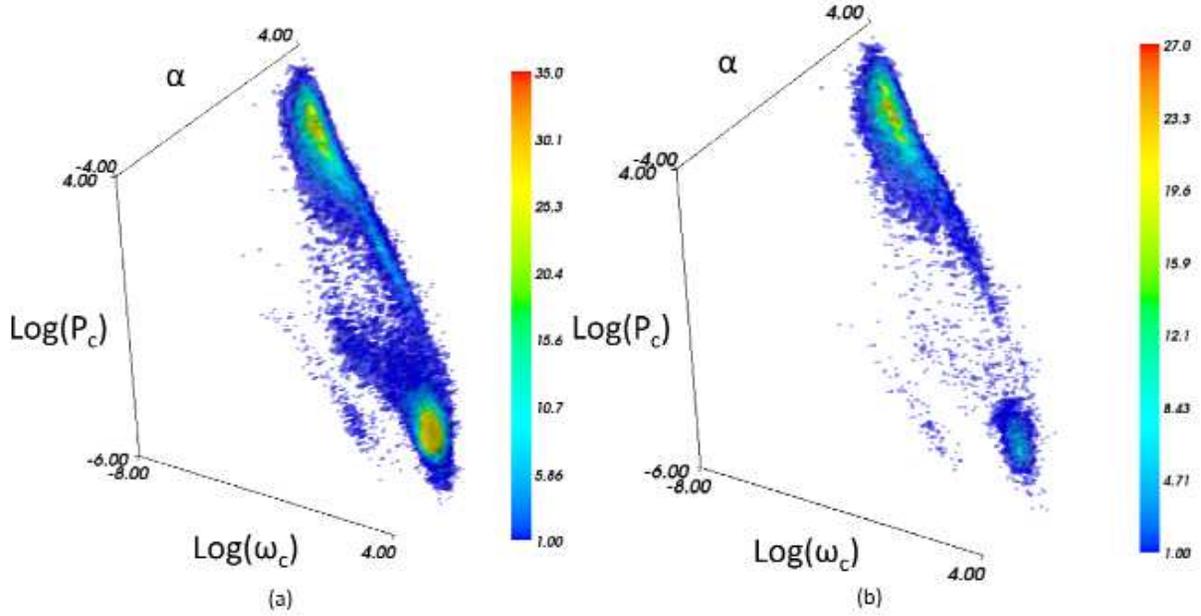}{0.9\textwidth}{}}
\caption{Distribution of derived parameters($\omega_{c}$, $\alpha$, $P_{c}$) on (a) all variable sources, 
and (b) only on those with spectroscopically identified QSOs and stars. They are the data based on 
those measured more than 40 times in $r$ band with errors less than 0.05 mag. 
\label{fig:3dcontour_allvar_qsostar}}
\end{figure}

We show the distribution of these three parameters($\alpha$, log $\omega_c$ and log $P_c$) on 67,507 variable sources 
by the 3-d contour plot in Figure \ref{fig:3dcontour_allvar_qsostar}(a), which indicates at least three clusters and 
two extensions in the distribution. 
The clusters are characterized by large $P_{c}$ and small $\omega_{c}$(Group1),  
small $P_{c}$ and large $\omega_{c}$(Group2), and small one neighboring to Group2 with smaller $\alpha$(Group3).  
There are two extensions, one is the narrow one connecting Group1 and Group2(Ext1), and 
another is extending to smaller $\alpha$ direction(Ext2).     

For checking the correspondence of the objects types (especially on QSOs and stars) to these structures  
in Figure \ref{fig:3dcontour_allvar_qsostar}(a), we plot the same figure only for the objects with 
spectroscopic confirmations as 'QSO' or 'STAR' in Figure \ref{fig:3dcontour_allvar_qsostar}(b). 
It is clear that the gap between Group1 and Group2 is clear in Figure \ref{fig:3dcontour_allvar_qsostar}(b), 
which is not visible in Figure \ref{fig:3dcontour_allvar_qsostar}(a). Additionally we can barely see Group3 
and Ext2, and can see Ext1 in the range of smaller $\omega$ range.   

It should be noted that the main contributors of these disappering structures are objects with small numbers 
of measurements, which suffer from the noisy inference of the parameters. 
We confirm that the objects labeled 'GALAXY' in the spectroscopic catalog of SDSS DR12 also belong to any of 
these structures. 

\begin{figure}[htb!]
\figurenum{8}
\plotone{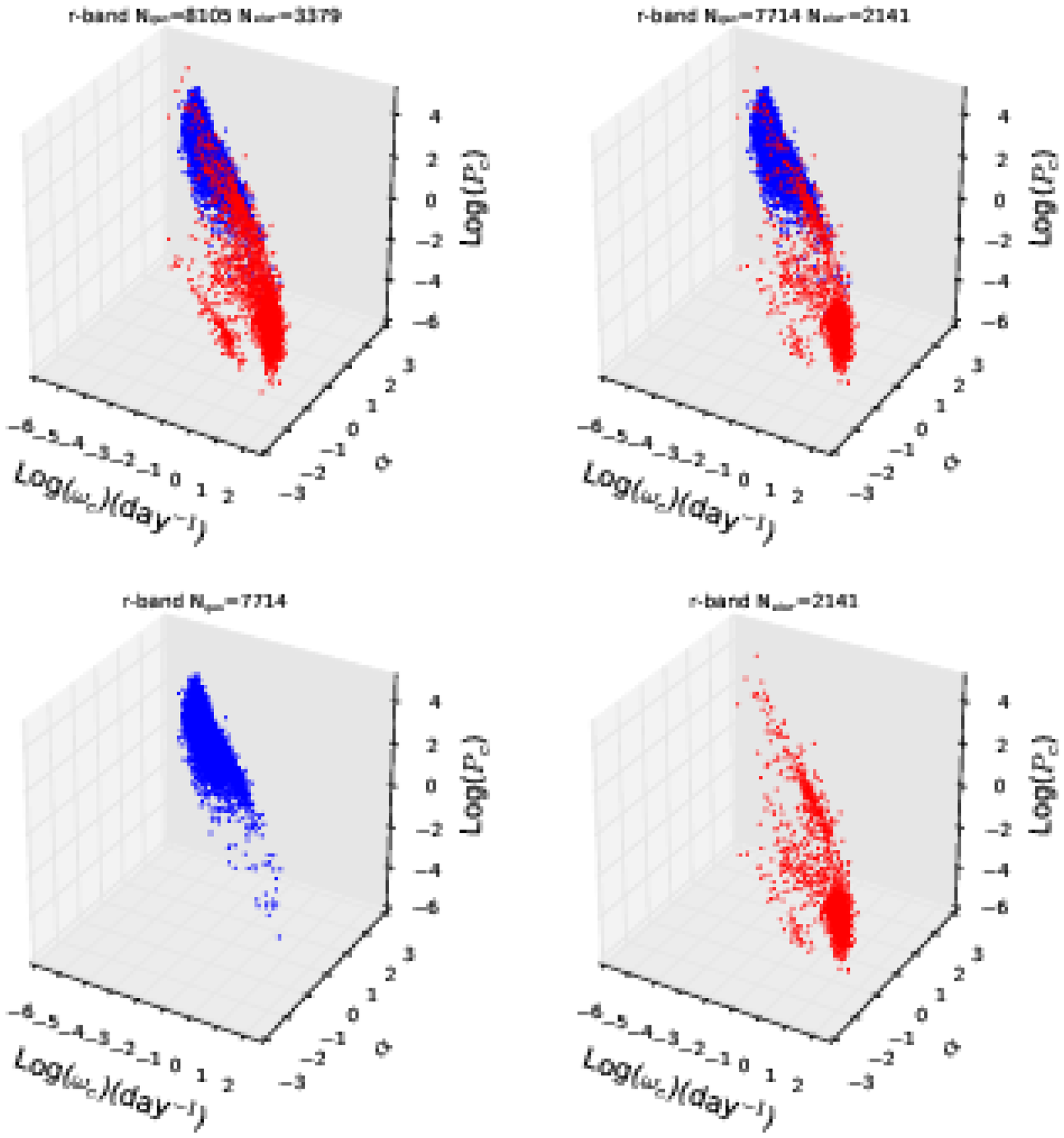}
\caption{Distribution of spectroscopically identified QSOs (blue) and stars(red) in derived 
parameters space ($\omega_{c}$, $\alpha$, $P_{c}$). Upper left figure shows all sample data 
(8,105 QSOs and 3,379 stars). Upper right figure is based on the 'good sample' data. 
There are 7,714 QSOs and 2,141 stars. Lower left and right figures show only QSOs and stars,respectively.  
\label{fig:3ddist_qsostar}}
\end{figure}

We also plot these three parameters of each spectroscopically confirmed QSOs and stars on the same parameter space 
in Figure \ref{fig:3ddist_qsostar}, to see the correlation between spectroscopic classification 
and variability features. 
These figures show that the majority of QSOs and stars belong to Group1 and Group2, respectively, 
and Group3 consists of stars. Stars are also the main contributors of two extensions.  

On the QSOs and stars in 'good sample', the effect caused by a small number of photometric measurements is reduced, 
and we can see more clearer separation of QSOs and stars in the parameter space.   
It means that distinguishing QSOs from stars is probably possible in very high identification rate, 
by using our model parameters, as probed by many previous similar works based only on optical variability 
(for examples, \citet{kozlowski2010a,macleod2010,macleod2011,butler2011,andrae2013}). 
The accuracy of the distinguishment should be higher in the case the number of photometric measurements increases. 

Another common clear feature is that the $\omega_{c}$ is larger the $P_{c}$ is smaller in QSOs(Group1).  
This means that the power of variability is higher the longer the damping time scale is. 
This is the same feature revealed by DRW model analysis\citep{macleod2011}.  

\section{Classification and Identification Rate of our Model}\label{sec:id_rate}
It is a well known fact that the optical variability information can be used as a powerful tool to 
distinguish a variable QSO population from other ones, such as variable stars, with high accuracy.
It is therefore very interesting to investigate the identification rate of variable QSO by the results based on 
our model calculation. As our method is based on the modeling of QSO/AGN variability, which comes from 
the long-term observational experience of many researchers, in the Fourier space, it is simply considered 
that the identification rate, especially completeness, may be higher than those by other models and calculations. 
It is because our model is customized to describe well the AGN-like variability, as well as DRW model, 
which is not suitable to other types of variability like periodically variable stars.     

We investigate the identification rate of the spectroscopically confirmed QSOs and stars, 
as performed in other studies for QSO classifications, because we can know the objects' types 
independently from variability. 
We compare the discrimination ability of the estimated model parameters or their subsets. 
In this case, we use $\log \omega_L, \log \omega_H, \alpha, \log \zeta, \log \omega_c, \log P_c$ 
and some of their subsets. 

As a classifier, we use linear Support Vector Machine(SVM). We use the linear classifier because
linear SVM scales well when the number of samples becomes large and it is easy to interpret the learned model.
We use the library of linear SVM in {\it{scikit-learn}} package\footnote{http://scikit-learn.org/} for our classification.  

\begin{figure}[htb!]
\figurenum{9}
\gridline{\fig{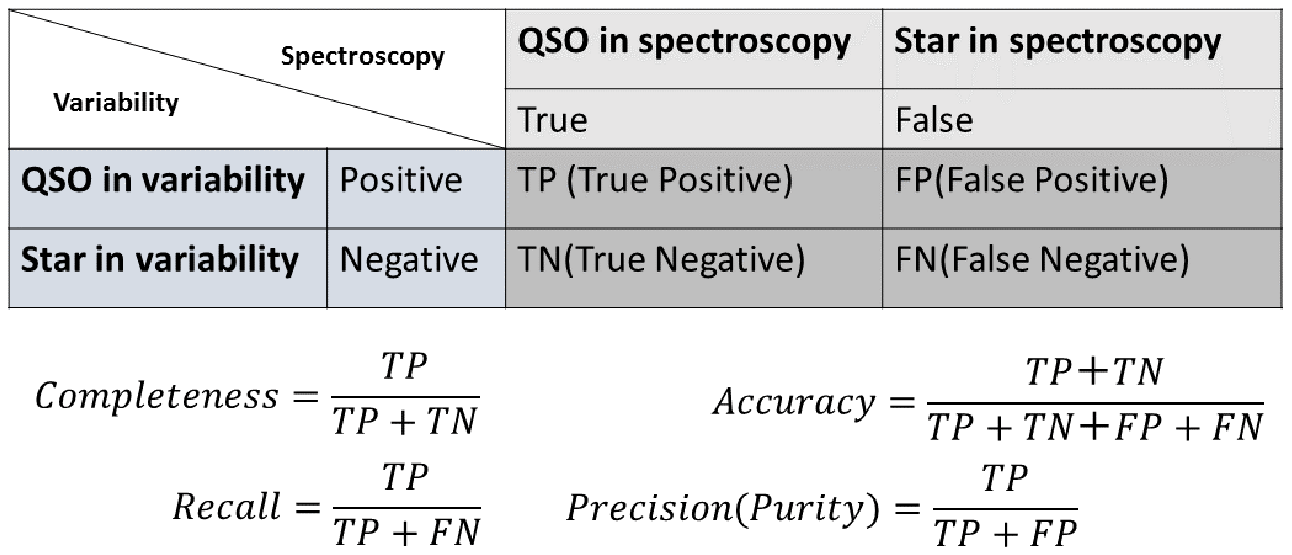}{0.6\textwidth}{}}
\caption{The definitions of values for describing identification rates used in our analysis. \label{fig:idrate_definition}}
\end{figure}

In each trial, we randomly choose 2,000 samples, half of which are QSOs and another half from stars, 
as test data and the rest as training data. We then decide the regularization parameter $C$ 
from $2^{-10}, 2^{-9}, ..., 2^{10}$ by applying 5-fold cross validation on the chosen training data. Then we learn
the classifier on the training data with determined $C$ and apply the learned model to the test data to get the results. 
We conducted 1,000 trials and evaluated the completeness, purity(precision), average of accuracy, and recall.
The definitions of these values in our analysis are shown in Figure \ref{fig:idrate_definition}. 

\begin{figure}[htb!]
\figurenum{10}
\gridline{\fig{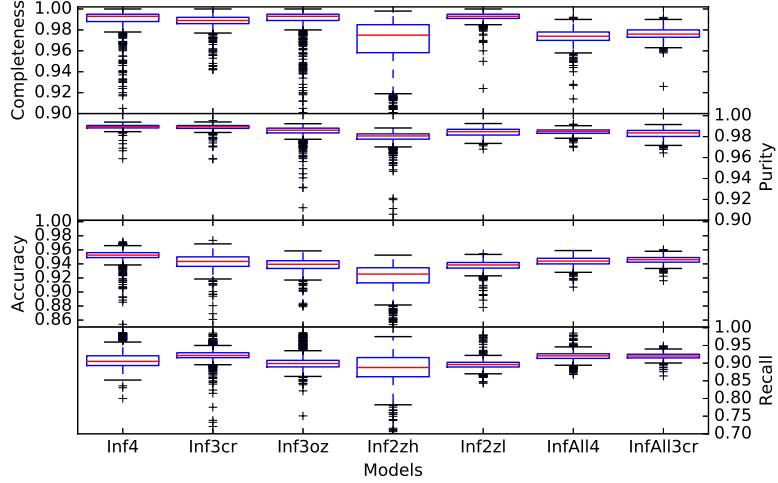}{0.6\textwidth}{}}
\caption{Box plot of the identification rates of our models. This figure shows 
the results on completeness, purity, accuracy, and average precision on $r$ band data, respectively. 
The box extends from the lower to upper quartile values of the data (blue lines), 
with red lines at the median. The whiskers extend from the box to show the range of 
the data, Q3 + 1.5$\times$IQR. IQR is inter-quartile value (Q3 - Q1), when Q1 and Q3 are 
the 1st and 3rd quartile values. Flier points are for those past the end of the whiskers. 
 \label{fig:idrate_infmodel}}
\end{figure}

In Figure \ref{fig:idrate_infmodel} we show the box-plot of the identification rates on our model 
based on 2,3,and 4 parameters space, and list the evaluated identification rates in Table \ref{tab:idrate_infmodel}. 
The identification rates shown here are based on the results by using the 'good sample'. 
Since we want to clarify the capability of optical variability to select QSOs, we constrain  
the discussions only to the results based on the 'good sample' hereafter if there is no notification.  

\floattable
\begin{deluxetable}{ll|rrrr}[htb!]
\tablecaption{Comparison of the Identification Rates on the sets of our model parameters\label{tab:idrate_infmodel}}
\tablehead{
\colhead{Abbreviation} & \colhead{Parameters} & \colhead{Completeness} & \colhead{Precision(Purity)} & \colhead{Accuracy} & \colhead{Recall} \\
}
%\colnumbers
\startdata
Inf4 & $\omega_L$,$\omega_H$,$\zeta$,$\alpha$& 0.993$^{+0.005}_{-0.002}$ & 0.990$^{+0.001}_{-0.001}$ & 0.956$^{+0.004}_{-0.004}$ & 0.905$^{+0.012}_{-0.016}$\\
Inf3cr & $\omega_{c}$,$P_{c}$,$\alpha$& 0.989$^{+0.003}_{-0.003}$ & 0.989$^{+0.002}_{-0.001}$ & 0.944$^{+0.007}_{-0.007}$ & 0.922$^{+0.006}_{-0.008}$\\
Inf3oz & $\omega_L$,$\omega_H$,$\zeta$ & 0.993$^{+0.005}_{-0.003}$ & 0.986$^{+0.003}_{-0.002}$ & 0.940$^{+0.006}_{-0.005}$ & 0.899$^{+0.010}_{-0.009}$\\
Inf2zh & $\omega_H$,$\zeta$ & 0.974$^{+0.016}_{-0.010}$ & 0.981$^{+0.003}_{-0.002}$ & 0.926$^{+0.013}_{-0.009}$ & 0.887$^{+0.026}_{-0.028}$\\
Inf2zl & $\omega_L$,$\zeta$ & 0.993$^{+0.002}_{-0.002}$ & 0.985$^{+0.003}_{-0.002}$ & 0.939$^{+0.005}_{-0.004}$ & 0.896$^{+0.007}_{-0.006}$\\
       &                    &                   &                   &                   &                  \\
InfAll4 & $\omega_L$,$\omega_H$,$\zeta$,$\alpha$& 0.974$^{+0.004}_{-0.004}$ & 0.985$^{+0.002}_{-0.002}$ & 0.944$^{+0.004}_{-0.004}$ & 0.921$^{+0.007}_{-0.006}$\\
InfAll3cr & $\omega_{c}$,$P_{c}$,$\alpha$& 0.976$^{+0.003}_{-0.004}$ & 0.984$^{+0.003}_{-0.002}$ & 0.946$^{+0.004}_{-0.003}$ & 0.920$^{+0.005}_{-0.005}$\\
\enddata
\end{deluxetable}

It is clear that the identification rates based on our optical variability modeling are very high, 
completeness is at about 97-99\% for all parameter sets, purities at 98-99\%, 93-95\% in accuracies, 
and 89-92\% in recall, respectively. There are very small differences among the identification rates by 
the different parameter sets. In completeness, the highest value is achieved by 'Inf4', 'Inf3oz', and 'Inf2zl' data set with 
99.3\%. 'Inf4' also provides the highest in purity(99.0\%), in accuracy(95.3\%), and 
'Inf3cr' in recall(92.2\%). It is noted that the rates are slightly decreased by adding the photometrically 
less measured sample (not 'good sample'), as represented by the data sets 'InfAll4' and 'InfAll3cr' in 
Table \ref{tab:idrate_infmodel}, which are based on the sample of all spectroscopically confirmed 
8,105 QSOs and 3,379 stars. 

\begin{figure}[htb!]
\figurenum{11}
\plotone{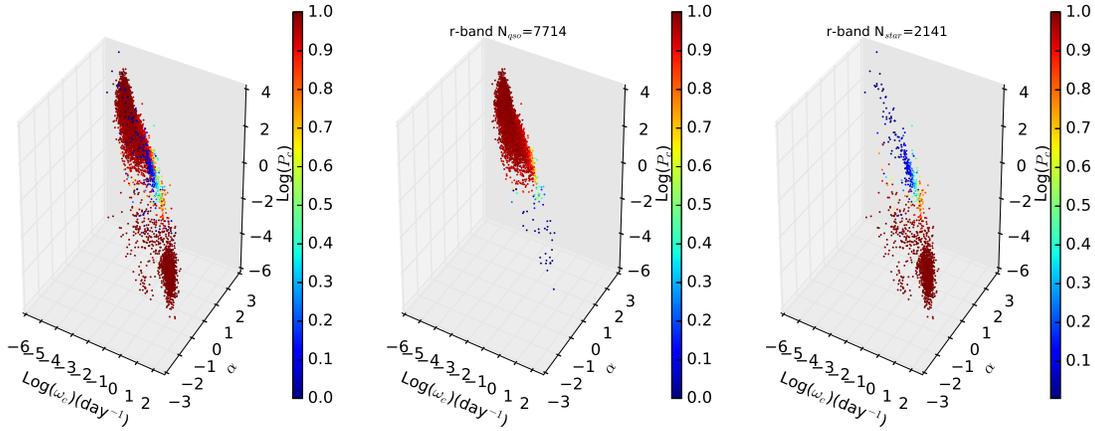}
\caption{The same as Figure \ref{fig:3ddist_qsostar}, but with their success rates in the 
classification using their optical variability with our model analysis. The left one is showing the 
distribution of spectroscopically confirmed QSOs and stars, and the middle and right plots are only 
for QSOs and stars, respectively. The colors show the success rates of each source. 
The color bar presents the success rates for each spectroscopically confirmed QSO and star.    
\label{fig:3ddist_success_rate}}
\end{figure}

By 1,000 trials of the classification using linear SVM, we can calculate the identification rates of our 
classification on each source, by counting up the numbers of successful classification on sources 
in the test sample. As the test data are randomly selected in each trial, all QSOs and stars in the 'good sample' 
are selected into test data homogeneously. Therefore we can calculate the identification rate 
for the whole sample without systematic biases. The range on the numbers of selection to test data  
is 120 to 190, with median value of $\sim$155, for QSOs and 520 to 600, with median value of $\sim$560, for stars.  
In Figure \ref{fig:3ddist_success_rate}, we show the distribution of QSOs and stars 
in three-dimensional parameter($\alpha$, log $\omega_c$, log $P_c$) space with their success rates, based on the results 
from our model('Inf4'). We can see the well-defined borders of QSOs and stars in our classification. 

\section{Discussions} \label{sec:discussions}
\subsection{Comparison with other models on identification rates}\label{subsec:id_rate_comp}
For confirming the performance of our model in QSO selection, we compare the identification rates 
(discrimination ability of 'QSO' and 'STAR') to several models. The models we use for the comparison are 
Butler's model\citep{butler2011}, Zu's single OU model\citep{zu2011, zu2013}, and Kelly's OU mixture 
models\citep{kelly2011}. 

\begin{figure}[htb!]
\figurenum{12}
\gridline{\fig{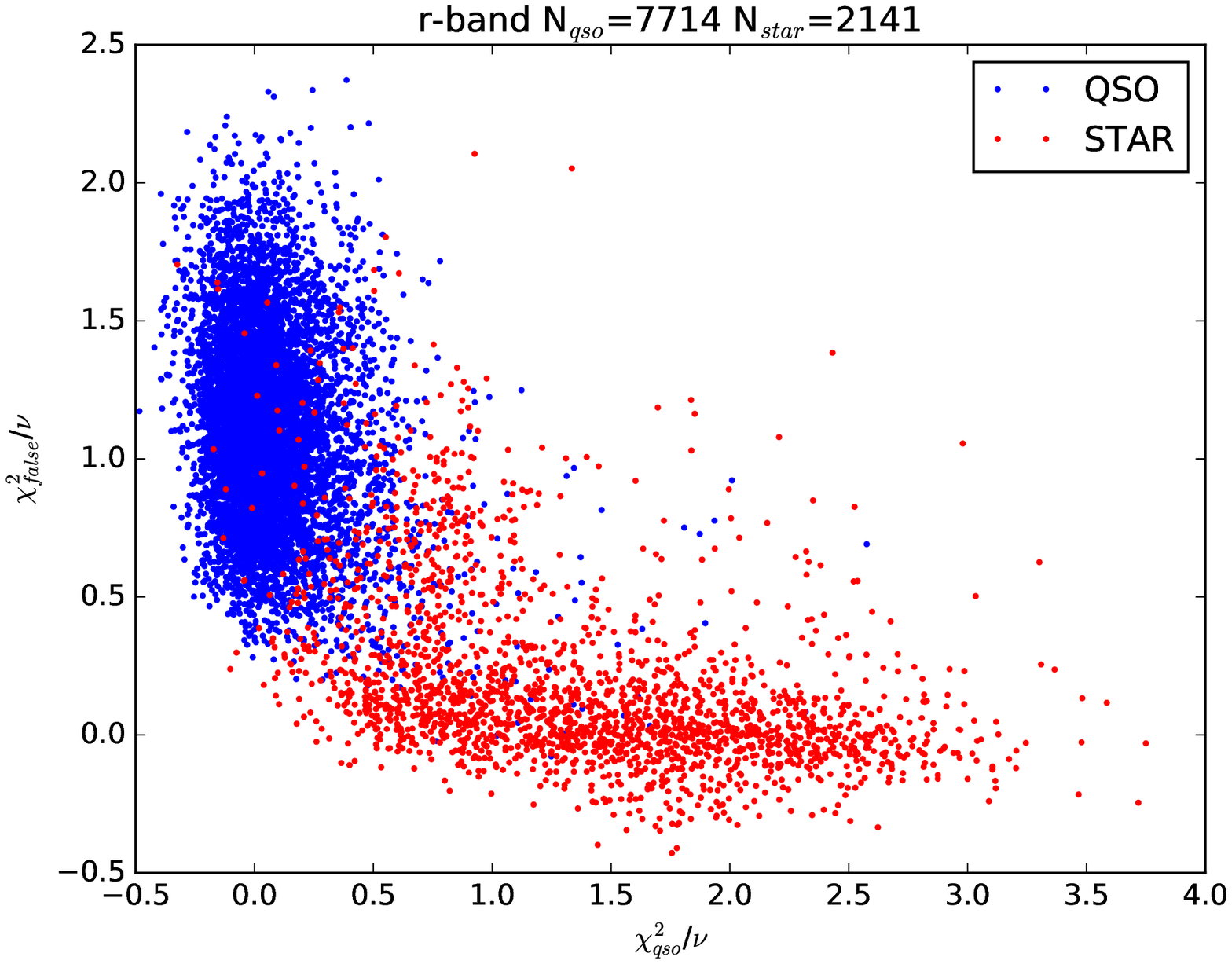}{0.45\textwidth}{(a)}
          \fig{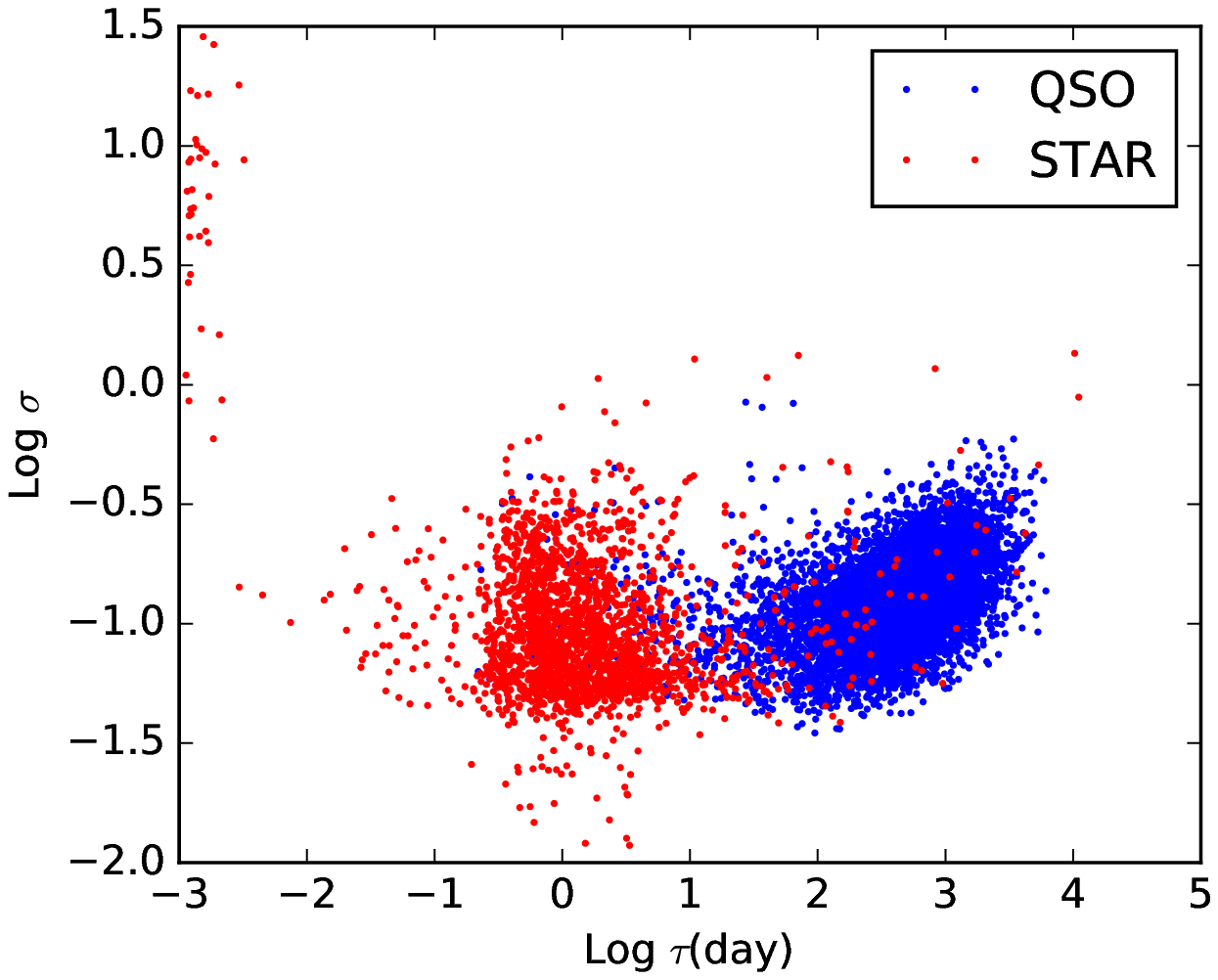}{0.45\textwidth}{(b)}}
\caption{(a)Distribution of model parameters of spectroscopically confirmed QSO(blue) and STAR(red) 
for the model of \citet{butler2011}. (b)That of QSOs and STARs parameters ($\tau$ and $\sigma$) for 
the model of \citet{zu2011}. 
The number of plotted QSO and STAR are 7,714 and 2,141, respectively. 
\label{fig:other_model_plot}}
\end{figure}

\citet{butler2011} use the DRW model by \citet{kelly2009} and implement the two parameters 
$\chi^{2}_{qso}$/$\nu$ and $\chi^{2}_{false}$/$\nu$ for well separating QSOs and stars in the parameter space, 
and it is believed to be one of the most powerful tools for 
selecting QSOs. We use their software released on their web page
\footnote{http://butler.lab.asu.edu/qso\_selection/index.html}.  
For computation using a single OU process, we use the {\rm{JAVELIN}} package developed and maintained by \citet{zu2011,zu2013}.  
Their program outputs $\tau$ and $\sigma$($\zeta$ in \ref{subsec:drwmodel}) and well separates QSOs and stars in two-dimensional parameter space. 
We develop the program for calculation based on Kelly's OU mixture model\citep{kelly2011}. We use the same MCMC program with 
our model calculation and we investigate in the cases of mixture number of OU process 
$M$ are equal to 2, 4, 8, 16, 32, 64 and 128. 
As the time consumed for the OU mixture model calculation is too long to calculate on all sample objects, 
we limit the calculation only to those with spectroscopic confirmation(11,908 objects). 
On the other hand we calculate 67,507 objects for Butler and single OU models. 
The distributions of QSOs and stars in the parameter spaces of each model are shown in Figure \ref{fig:other_model_plot}, 
and they are showing good performance on the separating QSOs from stars. 
\par

The classifications are performed in the same manner as for our model, and we derive the same values for 
estimating their discrimination abilities.  
Per each sample, we have $\log \tau, \log \sigma$ for single OU model, 
$\chi^{2}_{qso}$/$\nu$, $\chi^{2}_{false}$/$\nu$ for Butler's model and 
$\log \omega_L, \log \omega_H, \alpha, \log \zeta, \log \omega_c, \log P_c$ for Kelly's mixture model. 
We then apply the classifiers to the subsets of these parameters and determine whether the samples 
are QSOs or not.

\begin{figure}[htb!]
\figurenum{13}
\gridline{\fig{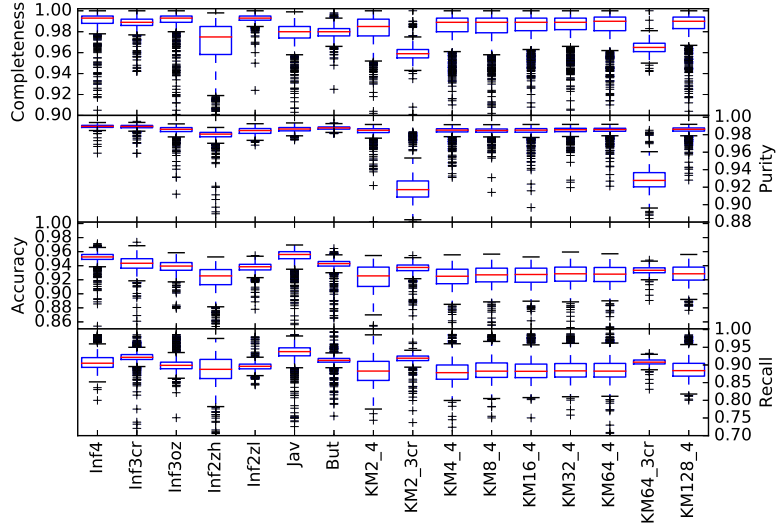}{0.6\textwidth}{}}
\caption{Box plot for comparisons of identification rates on the different models 
in the same manner as Figure \ref{fig:idrate_infmodel}. 
 \label{fig:idrate_all}}
\end{figure}

\floattable
\begin{deluxetable}{lll|rrrr}[htb!]
\tablecaption{Comparison of Identification Rates among models\label{tab:idrate_all}}
\tablehead{
\colhead{Model} & \colhead{Abbreviation} & \colhead{Parameters} & \colhead{Completeness} & \colhead{Precision(Purity)} & \colhead{Accuracy} & \colhead{Recall} \\
}
\startdata
Our Model  & Inf4 & $\omega_L$,$\omega_H$,$\zeta$,$\alpha$& 0.993$^{+0.005}_{-0.002}$ & 0.990$^{+0.001}_{-0.001}$ & 0.956$^{+0.004}_{-0.004}$ & 0.905$^{+0.012}_{-0.016}$\\
           & Inf3cr & $\omega_{c}$,$P_{c}$,$\alpha$& 0.989$^{+0.003}_{-0.003}$ & 0.989$^{+0.002}_{-0.001}$ & 0.944$^{+0.007}_{-0.007}$ & 0.922$^{+0.006}_{-0.008}$\\
           & Inf3oz & $\omega_L$,$\omega_H$,$\zeta$ & 0.993$^{+0.005}_{-0.003}$ & 0.986$^{+0.003}_{-0.002}$ & 0.940$^{+0.006}_{-0.005}$ & 0.899$^{+0.010}_{-0.009}$\\
           & Inf2zh & $\omega_H$,$\zeta$ & 0.974$^{+0.016}_{-0.010}$ & 0.981$^{+0.003}_{-0.002}$ & 0.926$^{+0.013}_{-0.009}$ & 0.887$^{+0.026}_{-0.028}$\\
           & Inf2zl & $\omega_L$,$\zeta$ & 0.993$^{+0.002}_{-0.002}$ & 0.985$^{+0.003}_{-0.002}$ & 0.939$^{+0.005}_{-0.004}$ & 0.896$^{+0.007}_{-0.006}$\\
\hline
JAVELIN    & Jav  & $\tau$,$\sigma$ & 0.980$^{+0.007}_{-0.005}$ & 0.986$^{+0.002}_{-0.002}$ & 0.956$^{+0.006}_{-0.004}$ & 0.938$^{+0.012}_{-0.011}$\\
BUTLER     & But & $\chi^{2}_{qso}$/$\nu$, $\chi^{2}_{false}$/$\nu$ & 0.980$^{+0.004}_{-0.003}$ & 0.988$^{+0.001}_{-0.001}$ & 0.943$^{+0.004}_{-0.003}$ & 0.913$^{+0.005}_{-0.006}$\\
Mix(M=2)   & KM2\_4 & $\omega_L$,$\omega_H$,$\zeta$,$\alpha$ & 0.985$^{+0.010}_{-0.007}$ & 0.985$^{+0.003}_{-0.002}$ & 0.926$^{+0.015}_{-0.013}$ & 0.882$^{+0.026}_{-0.027}$\\
Mix(M=4)   & KM4\_4 & $\omega_L$,$\omega_H$,$\zeta$,$\alpha$ & 0.989$^{+0.010}_{-0.004}$ & 0.985$^{+0.002}_{-0.002}$ & 0.925$^{+0.011}_{-0.010}$ & 0.878$^{+0.019}_{-0.022}$\\
Mix(M=8)   & KM8\_4 & $\omega_L$,$\omega_H$,$\zeta$,$\alpha$ & 0.989$^{+0.010}_{-0.004}$ & 0.985$^{+0.002}_{-0.002}$ & 0.927$^{+0.010}_{-0.010}$ & 0.882$^{+0.017}_{-0.023}$\\
Mix(M=16)  & KM16\_4 & $\omega_L$,$\omega_H$,$\zeta$,$\alpha$ & 0.989$^{+0.009}_{-0.004}$ & 0.985$^{+0.002}_{-0.002}$ & 0.928$^{+0.011}_{-0.009}$ & 0.882$^{+0.018}_{-0.020}$\\
Mix(M=32)  & KM32\_4 & $\omega_L$,$\omega_H$,$\zeta$,$\alpha$ & 0.989$^{+0.008}_{-0.004}$ & 0.986$^{+0.002}_{-0.002}$ & 0.929$^{+0.011}_{-0.010}$ & 0.883$^{+0.017}_{-0.021}$\\
Mix(M=64)  & KM64\_4 & $\omega_L$,$\omega_H$,$\zeta$,$\alpha$ & 0.990$^{+0.010}_{-0.004}$ & 0.986$^{+0.002}_{-0.002}$ & 0.928$^{+0.011}_{-0.010}$ & 0.882$^{+0.017}_{-0.022}$\\
Mix(M=128) & KM128\_4 & $\omega_L$,$\omega_H$,$\zeta$,$\alpha$ & 0.990$^{+0.008}_{-0.004}$ & 0.986$^{+0.002}_{-0.002}$ & 0.929$^{+0.009}_{-0.010}$ & 0.884$^{+0.016}_{-0.021}$\\ 
\hline
Mix(M=2)   & KM2\_3cr & $\omega_c$,$P_c$,$\alpha$ & 0.959$^{+0.004}_{-0.004}$ & 0.917$^{+0.008}_{-0.010}$ & 0.938$^{+0.005}_{-0.004}$ & 0.919$^{+0.007}_{-0.006}$\\
Mix(M=64)  & KM64\_3cr & $\omega_c$,$P_c$,$\alpha$ & 0.965$^{+0.004}_{-0.004}$ & 0.928$^{+0.007}_{-0.009}$ & 0.934$^{+0.004}_{-0.004}$ & 0.908$^{+0.005}_{-0.006}$\\
\enddata
\end{deluxetable}

In Figure \ref{fig:idrate_all}, we show the box-plots for the comparison of identification rates 
among these models, and list their identification rate values in Table \ref{tab:idrate_all}.  
It is clear that our model has the highest rate in completeness, especially 99.3\% in the cases of 'Inf4', 'Inf3oz' and 'Inf2zl', 
and we've confirmed those of Kelly's model are also high compared to other models.  
It is very plausible that ours and Kelly's models are more flexible model for describing QSO/AGN variability. 
The significance of the difference in completeness from those of other models such as \citet{butler2011} and \citet{zu2011} models 
is verified by the pairwise multiple comparison test\citep{demsar2006}
\footnote{http://finzi.psych.upenn.edu/library/PMCMR/html/PMCMR-package.html}. 
In terms of recall, single OU model by \citet{zu2011} shows the highest score, but the score of our model is higher than 
those of other models. 

It should be noted that the improvement of identification rates of QSOs, especially in relation to completeness, by 
our model compared to others comes from the discrimination ability on the objects at the edge of the Group1. 

\begin{figure}[htb!]
\figurenum{14}
\plottwo{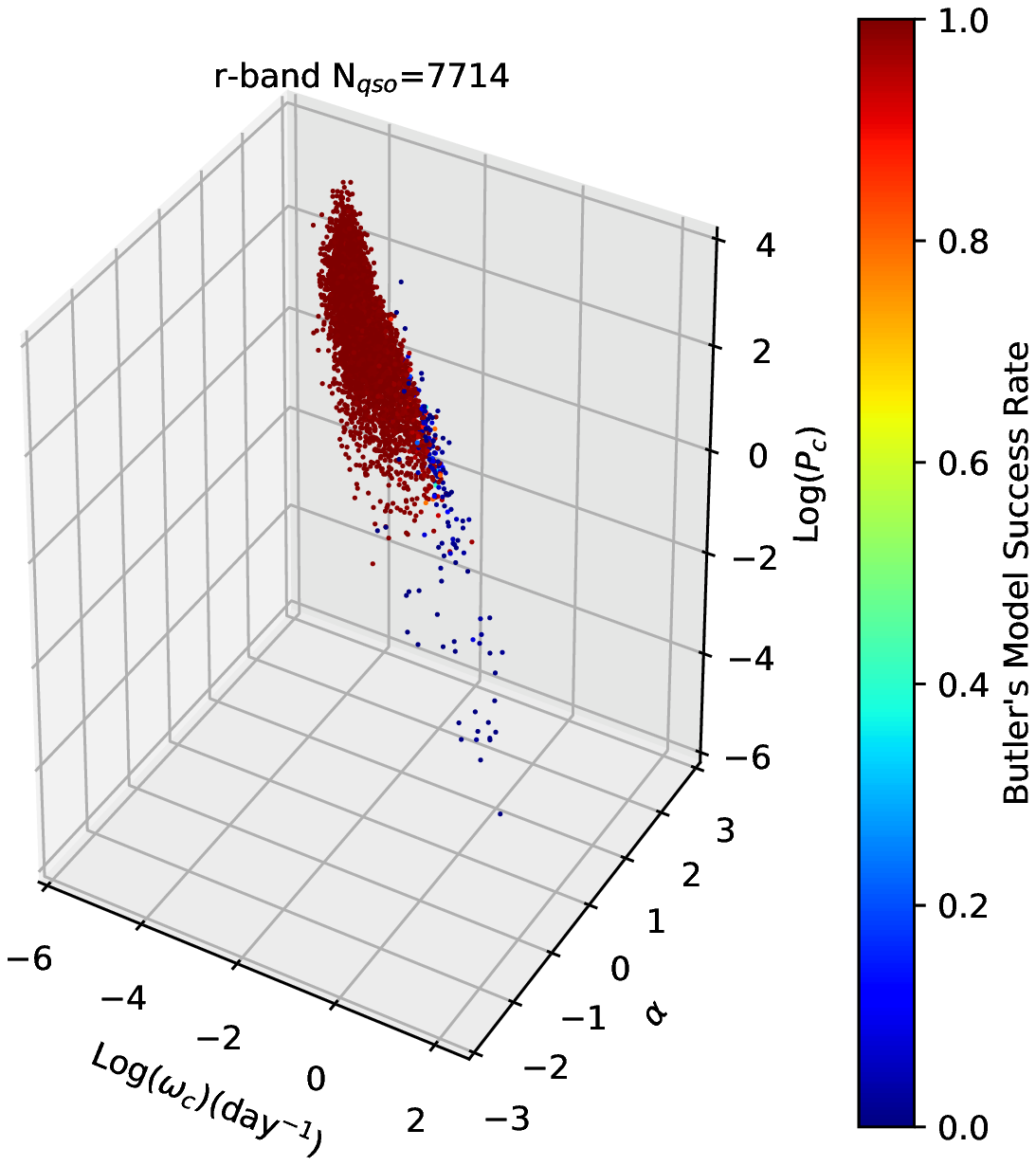}{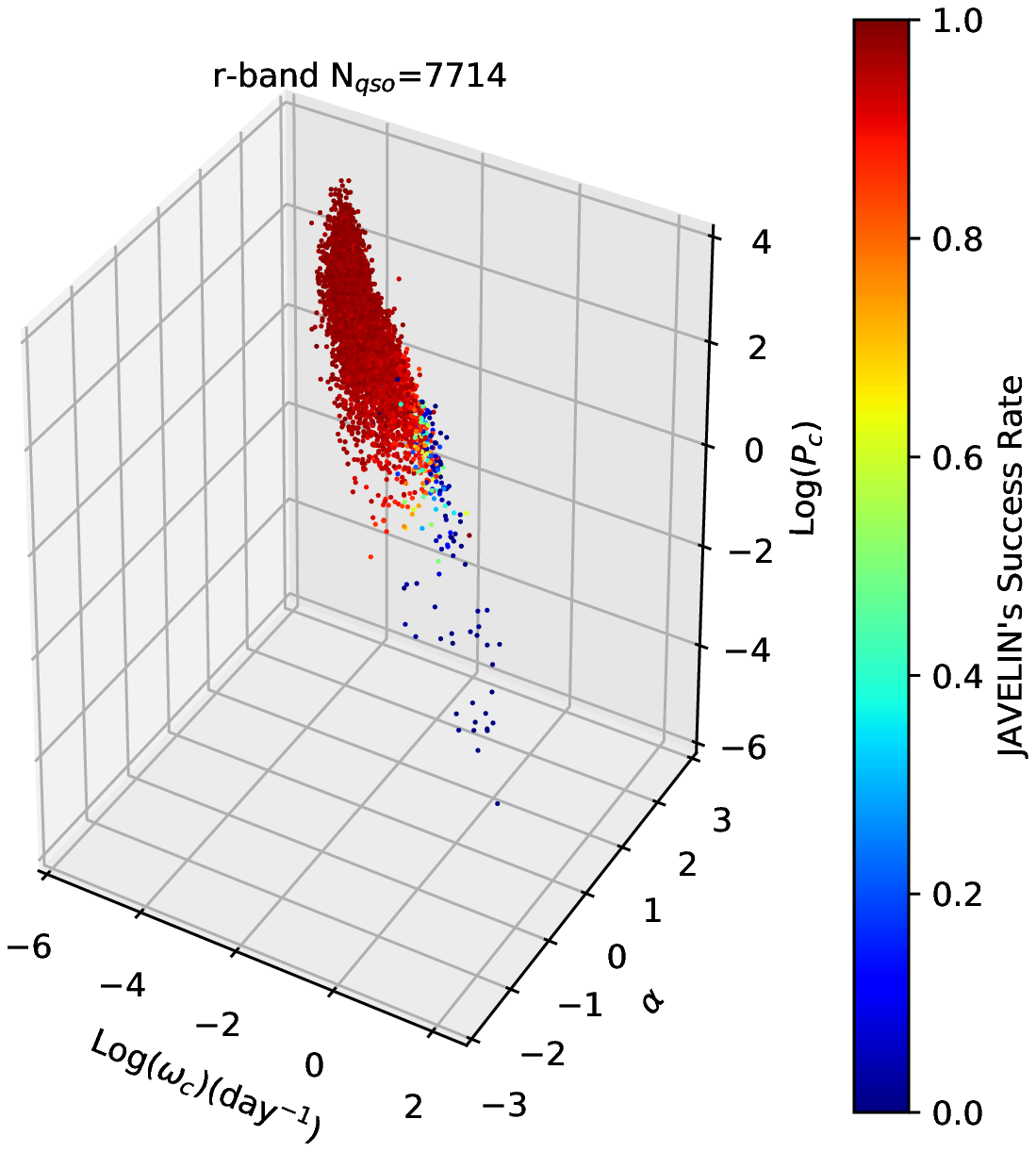}
\caption{The same one with middle panel of Figure \ref{fig:3ddist_qsostar}, but with the success 
rates with \citet{butler2011}'s model(left) and {\rm{JAVELIN}}(right). 
 \label{fig:3ddist_success_rate_other_model_qso}}
\end{figure}

In Figure \ref{fig:3ddist_success_rate_other_model_qso}, we show the three-dimensional distribution of our model 
parameters for spectroscopically confirmed QSOs with the success rates by \citet{butler2011}'s model  
and {\rm{JAVELIN}}. By comparing the success rates distribution by our model, shown in the middle of 
Figure \ref{fig:3ddist_success_rate}, it is clear that most objects in the group(Group1) are classified successfully 
as 'QSO' in our model, although some misclassifications can be seen in the results by other models. 
It is therefore reasonable to explain this improvement is caused by the introduction of the mixed OU processes in our model. 

We should, however, be cautious regarding the bad success rate on 'Inf2zh', which is classified based on 
$\omega_H$ and $\zeta$, instead of very high performance on 'Inf2zl'. 
It implies that QSOs with small damping time scale are indistinguishable from stars on the parameter plane 
defined by $\omega_H$ and $\zeta$. 
It should be noted that the larger number of mixtures of OU processes provide better success rates in completeness, 
although saturated around $M=8$. 

\begin{figure}[htb!]
\figurenum{15}
\gridline{\fig{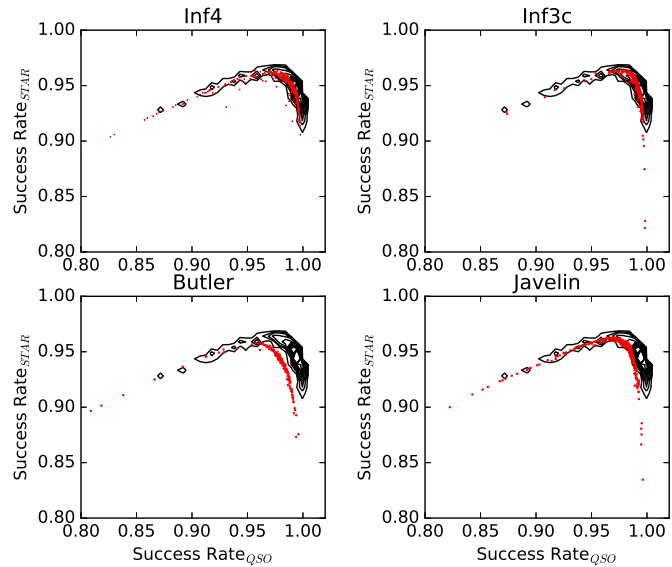}{0.6\textwidth}{}}
\caption{Comparison of identification rates of QSOs and stars for 1000 sets of coefficients in our SVM analysis.
Contours in each panel show the distribution for the 'Inf4' model, and the red points represent the results 
of 1000 trials for each model. They are for Inf4, Inf3cr, Butler's model, and Javelin(DRW model), respectively, 
in the order of upper left, upper right, lower left, and lower right.    
\label{fig:comparison_qsostar_idrate}}
\end{figure}

In Figure \ref{fig:comparison_qsostar_idrate}, we show the distribution of success rates of QSOs and stars in our 
1,000 trials of classification using linear SVM. We compare the performance of the 'Inf4' model 
to others ('Inf3cr', 'JAVELIN', and 'Butler') in this figure. The identification rates are 
computed for the whole 'good sample', on 9,855 sources. 
In this figure we can see the clear advantage the 'Inf4' model has over the 'Butler' model in 
QSOs and stars selection, and over 'JAVELIN' in QSO selection in the high identification-rate region. 
On the other hand, we cannot see any significant difference from 'Inf3cr'. 

As we use linear SVM for our classification, the hyperplane separating the parameter zones for QSOs and stars 
can be expressed as follows; 

\begin{equation}\label{eq:hyperplane_inf4}
A \alpha + B {\rm{Log}}_{10}(\omega_{L}) + C {\rm{Log}}_{10}(\omega_{H}) + D {\rm{Log}}_{10}(\zeta) + E \left\{ \begin{array}{ccl}
                                                                                         < & 0 & (QSO) \\
                                                                                         > & 0 & (star)
                                                                                       \end{array} \right.
\end{equation}

As the number fraction of QSOs to stars depends on sky region and/or depth of the imaging data/catalog, 
it is convenient to use different hyperplanes for selecting QSOs from the variable object samples, 
as the 'total' success rates, defined as the weighted mean of success rates for QSO and star selections. 
The information may be one of the important issues to decide the strategy for QSO selection using the given data set.   
For example, we should use conservative criteria to eliminate contamination by stars, if the purity of QSO selection 
is the highest priority on the selection. 

%% Table for SVM coefficients 
\floattable
\begin{deluxetable}{rrrrrrrrr}[htb!]
\tablecaption{Coefficients of SVM hyperplane to separate QSOs and stars with best weighted identification rate for given number fractions of QSOs and stars.\label{tab:svmcoeffs}}
\tablehead{
\colhead{N$_{QSO}$/N$_{star}$} & \colhead{A} & \colhead{B} & \colhead{C} & \colhead{D} & \colhead{E} & \colhead{R$_{total}$}$^{a}$ & R$_{QSO}$$^{b}$ & R$_{star}$$^{b}$ \\
}
%\colnumbers
\startdata
0.1 & -0.728   & 0.063  & 0.227 & -0.069  & 1.297 & 0.971 & 0.870 & 0.981\\  
0.5 & -0.664   & 0.139  & 0.211 & -0.019  & 1.209 & 0.965 & 0.957 & 0.970\\
1.0 & -0.775   & 0.146  & 0.201 &  0.151  & 1.560 & 0.964 & 0.966 & 0.961\\
1.2 & -0.695   & 0.119  & 0.255 & -0.073  & 0.945 & 0.965 & 0.977 & 0.950\\
1.5 & -0.695   & 0.119  & 0.255 & -0.073  & 0.945 & 0.966 & 0.977 & 0.950\\
2.0 & -0.774   & 0.189  & 0.222 &  0.172  & 1.551 & 0.968 & 0.983 & 0.939\\
2.5 & -0.768   & 0.210  & 0.250 &  0.111  & 1.467 & 0.971 & 0.987 & 0.931\\
3.0 & -0.086   & 0.425  & 0.395 & -0.410  & 0.949 & 0.972 & 0.988 & 0.924\\
3.0 & -0.768   & 0.210  & 0.250 &  0.111  & 1.467 & 0.973 & 0.987 & 0.931\\
5.0 & -0.575   & 0.285  & 0.253 & -0.101  & 1.159 & 0.978 & 0.991 & 0.913\\
\enddata
\tablenotetext{a}{Total identification(success) rates are calculated by weighted mean of those for QSO(R$_{QSO}$) and stars(R$_{star}$).}
\tablenotetext{b}{QSO(R$_{QSO}$) and stars(R$_{star}$) are values of completeness.}
\end{deluxetable}

We show some sets of hyperplane coefficients for separating QSOs and stars in four parameters 
space with maximum total success rates (weighted mean of QSOs and stars success rates) in 
Table \ref{tab:svmcoeffs} with an assumed number fraction of QSOs to stars in the sample. 
Note that in our photometric sample, the fraction varies from about 0.1 to 2.2. 
For example, in case we apply the coefficients for N$_{QSO}$/N$_{star}$=1.5, 
which is the rough average of the data set, we select 9,077 QSO candidates out of 
23,677 variable sources, which have more than 40 measurements with errors less than 0.05 magnitudes, 
based only on photometric data. In this case it will provide additional $\sim$1,300 QSO candidates  
to the current spectroscopically confirmed QSOs in the data set. 

\subsection{Failure rates and their meanings}\label{subsec:failure_rate}
In our analysis, we identified 83 spectroscopically confirmed QSOs with the 'false' classification 
by their optical variability. The 'false' means the success rate is lower than 50\% in our analysis. 
It is about one percent of the spectroscopically confirmed QSOs and it is interesting to know the reason for this failure 
in the classification.  
We visually inspect all the spectra of 83 QSOs, and confirm that 79 objects can be reasonably classified as QSO/AGN. 
The remaining four objects are three stars and one object with failure in taking the spectrum(fiber allocation might have failed).  
It should be noted that the misclassification occurs on the common objects by the analysis based on other models, 
such as Butler's model, single-OU and Kelly's mixture model, and it is not a characteristic feature of our model. 
It means that their variability can not be distinguished from those of stars. 

There are several possibilities for explaining the failure of classification. 
The first is the effect caused by the irregularly taken time series data, that hides or dulls the 
emergence of the QSO/AGN-like variability features in their light curves. It may be coming from the limitation 
of MCMC procedure for producing the light curves for our stochastic model analysis. 
It is however difficult to consider this a plausible reason, as the most of objects are successfully classified as 'QSO'   
in spite of being under the conditions of very similar cadence, as described in \ref{subsec:speccat}. 

The second possibility is that this discrepancy comes from the time difference between photometric and spectroscopic 
observations. As we are based on the SDSS DR12 data on the spectroscopic data, although the photometric data is coming 
from DR7, there is a possibility of a change of physical condition in these objects. 
We investigate the MJDs of the data taking on photometry and spectroscopy, and confirm that 40 and 43 spectral data are 
taken in and out of the duration on the photometric data. 
Therefore, it is difficult to consider the time difference of photometry and spectroscopy as the main reason for the discrepancy. 

The third and most plausible is that the objects are rare type QSOs/AGNs whose variability features 
differ from those of most QSOs. There may be the signatures of (quasi-) periodic features 
in their light curves and/or cannot be well described by MCMC procedure as mentioned above. 

\begin{figure}[htb!]
\figurenum{16}
\gridline{\fig{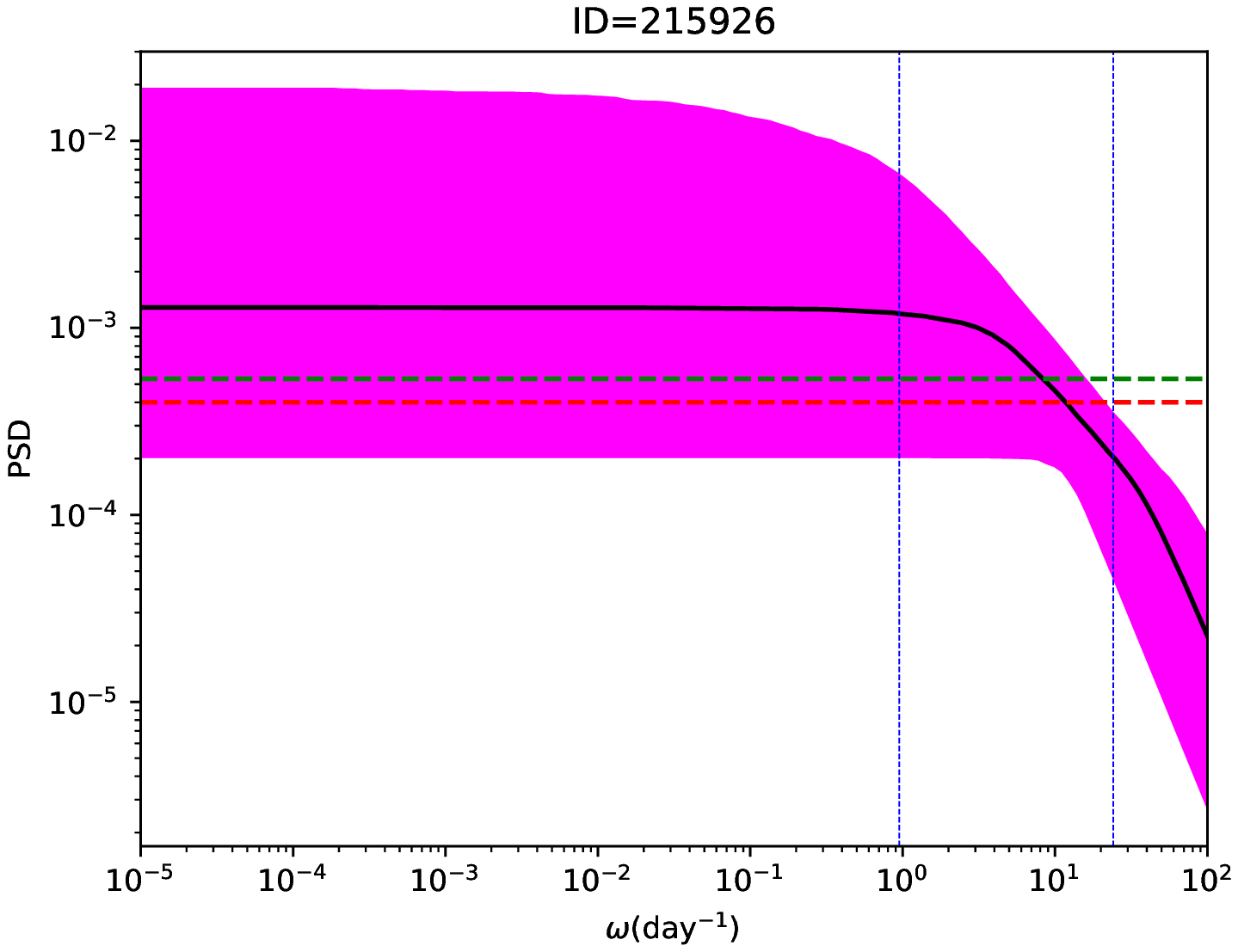}{0.4\textwidth}{(a)}
          \fig{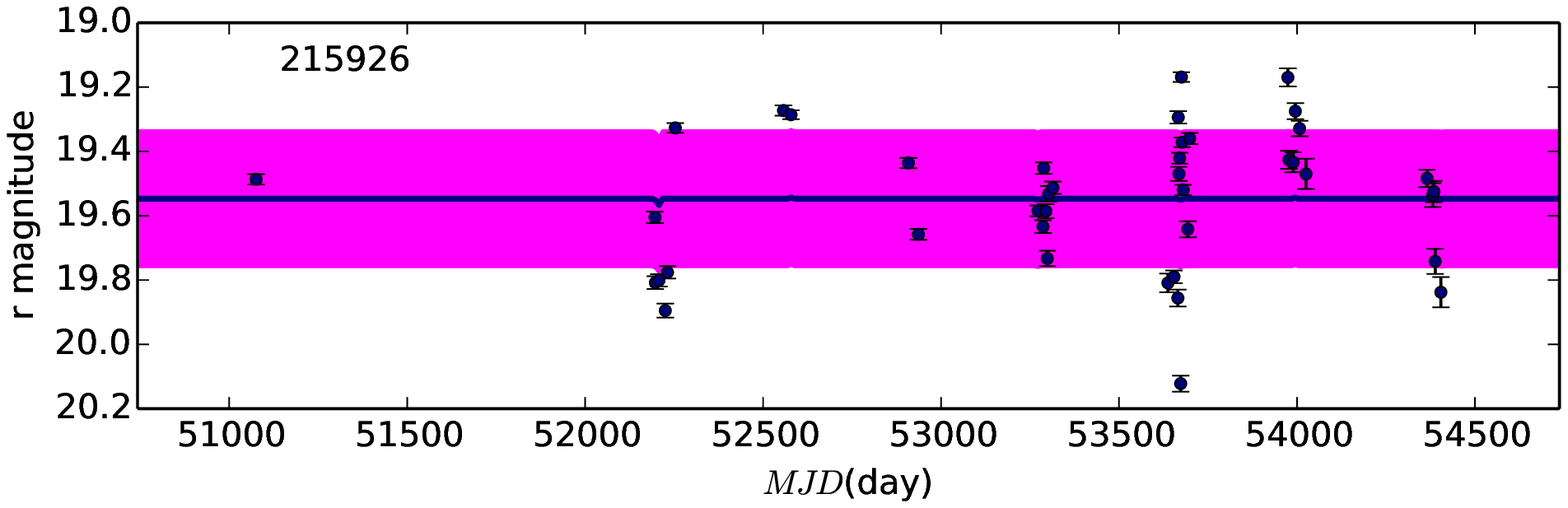}{0.4\textwidth}{(b)}}
\caption{(a) A PSD of a spectroscopically confirmed QSO categorized in Group2, which can not be distinguished 
from those of stars. (b) Light curve of the same object in $r$ band magnitudes.  
 \label{fig:spqso_varstar}}
\end{figure}

As shown in the left panel of Figure \ref{fig:spqso_varstar}(a), objects classified as not 'QSO' 
(categorized in Group2 as described in \ref{subsec:psd}) have single-bending PSDs 
with very large $\omega$ at the bending points, which is very similar to those for stars in our analysis. 
We also show the light curve of the same object in Figure \ref{fig:spqso_varstar}(b), and it is clear that the variation in short 
time scale is so large and we cannot see the damping of brightness, which is seen in typical QSOs, as shown 
in Figure \ref{fig:infmodel_analysis_qso}. This implies the rarity (they contribute only one percent of spectroscopically confirmed QSOs) 
of this QSO in terms of optical variability, and there might be a different physical process controlling the variation of brightness, 
at least in a short time scale.  

We cannot see any significant difference in the spectra of the QSOs with 'false' classification, 
compared to the typical variable QSOs, and it is plausible to consider the discrepancies as coming from 
the physical processes which are effective to optical variability although not to optical spectral features. 

As the fractions of QSOs and stars spectroscopically misidentified are so small
(4/83 for QSOs and 16/199 for stars, respectively), it is clear that the results 
of identification rates discussed in previous sections are not affected seriously.   

\subsection{Features of our model and Interpretation of QSOs' PSD parameters}\label{subsec:agn_psd_param}
We compare the goodness of the fit of our model to single OU process(DRW) model by using 
AIC and BIC(Bayesian information criterion)\citep{schwarz1978}. 

AIC is derived by the following equation,  

\begin{equation}\label{eq:aic}
 AIC = -2 \ln(L) + 2k,
\end{equation}
where $L$ is the maximum value of likelihood function and $k$ is 
the number of estimated parameters in the model. 
In the case of our model, $k$ is 4, and $k$ is 2 for DRW model.  
BIC is derived by the Equation \ref{eq:bic} as described in \ref{subsec:fitlc_model}. 

\begin{figure}[htb!]
\figurenum{17}
\gridline{\fig{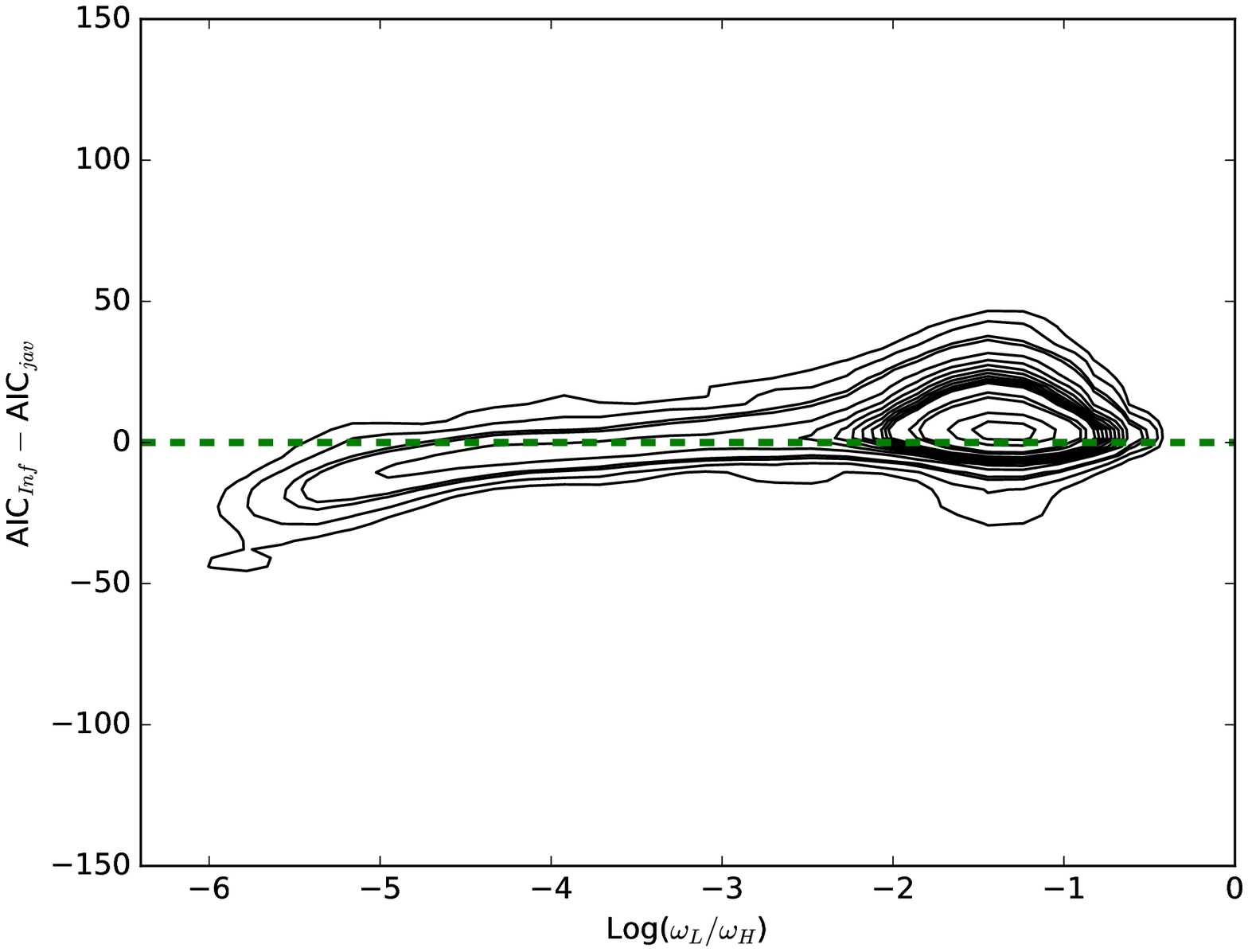}{0.45\textwidth}{(a)}
          \fig{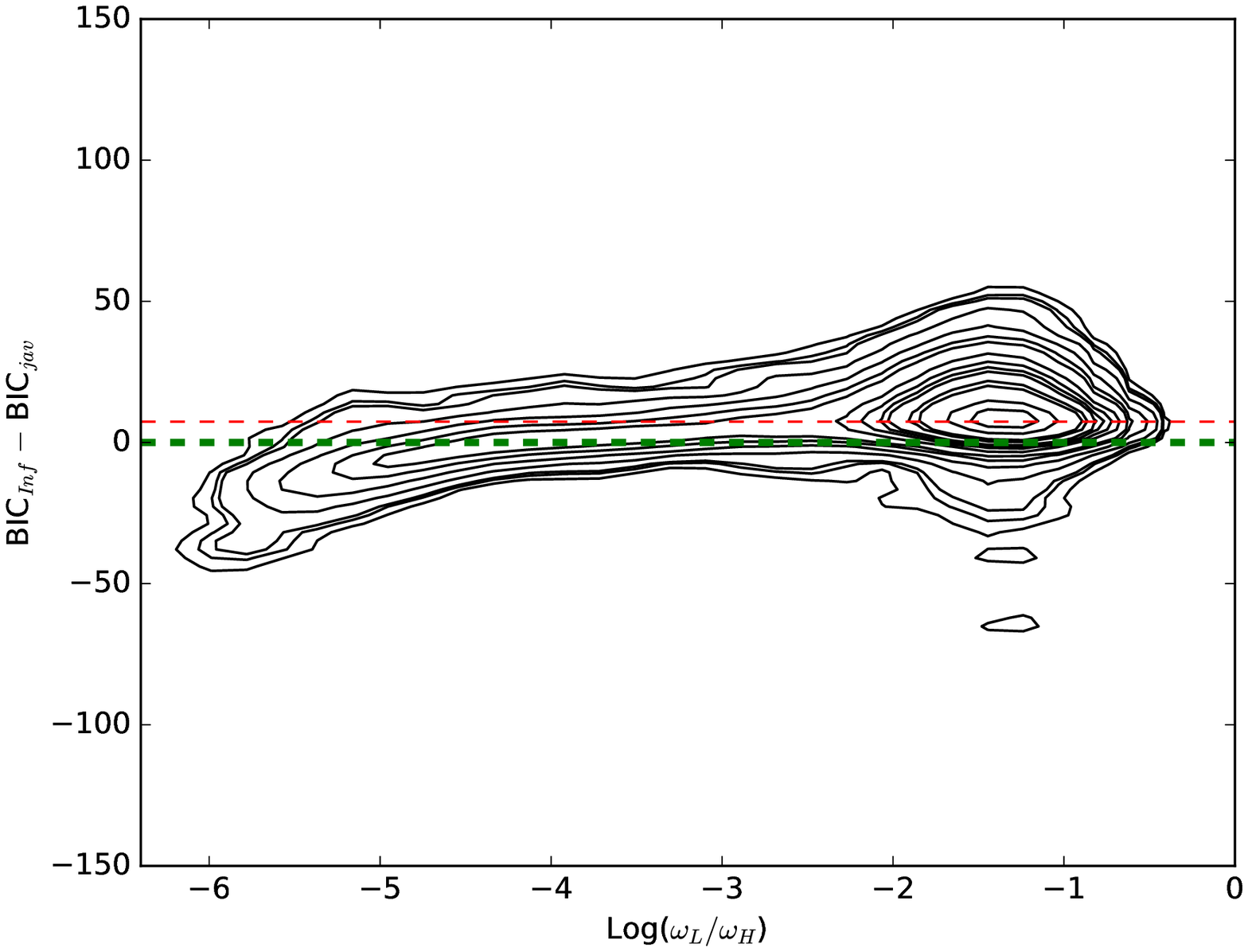}{0.45\textwidth}{(b)}}
\caption{(a)The dependencies of the differences of AIC values between those for our model and DRW model, to the ratio of 
$\omega_L$ and $\omega_H$ in our model. The horizontal green dashed line shows the border for equal AIC values, 
which means the two models have the same rank. 
(b)The same figure for BIC values. The horizontal green line means the same as \ref{fig:aicbic_for_qso}(a), and 
red dashed line show the case when the likelihoods are the same in the case number of measuremets is 40.  
 \label{fig:aicbic_for_qso}}
\end{figure}

In Figure \ref{fig:aicbic_for_qso}, we show the comparison of AICs and BICs for QSOs in the 'good sample'. 
It is clear that our model works better for QSOs with large differences between $\omega_L$ and $\omega_H$, 
which correspond to variability different from DRW, since the smaller AIC/BIC values mean better fitness. 

The trends seen about AICs are the same for BICs, that our model works better for QSOs 
with larger discrepancies between $\omega_L$ and $\omega_H$, for example $\log \frac{\omega_L}{\omega_H} \leq -3$.  

It is therefore reasonable to think our model provides better fit than those based on DRW model 
for QSOs with larger differences between $\omega_L$ and $\omega_H$. 
It should be noted that many QSOs, whose variability can be well described by DRW model, show worse fitting results 
in our model. However, this causes no significant effect for selection of QSOs/AGNs from variable stars. 
The differences are only coming from the number of model parameters, and the likelihood values are equivalent 
about both models. 
\par
We also note here that the AIC in its original form is strictly only valid asymptotically, 
and the correction to AIC for finite sample sizes suggested by \citet{sugiura1978} and \citet{hurvich1989}, 
commonly denoted as c-AIC, may be used for the discussions.   
We also check the c-AIC and confirm that there is no difference from the result by using AIC.    

The PSDs for the QSOs/AGNs with smaller AIC/BIC in our model show the power law like features, which have 
flatter slopes than $\beta$=-2, when we express the PSD $ \propto \omega^{\beta}$. In most cases they have 
slopes $\beta \sim$-1.8 or so. 
They have very small $\omega_L$($\log \omega_L \sim$-4) and large $\omega_H$ ($\log \omega_L \sim$1-2) as 
shown 'PowerLaw' type variability in Figure \ref{fig:covariance_function}(b). 
In our model such a variability can be expressed well than DRW model by selecting large difference in 
$\omega_L$ and $\omega_H$.  

On the other hand, most QSO have small differences between $\omega_L$ and $\omega_H$, typically 
$\log \omega_L/\omega_H \sim$ -1. In our model there is no constraint about the values of 
$\omega_L$ and $\omega_H$, except for  $\omega_L \leq \omega_H$. 
One of the reason for this small diffference is our method for estimating model parameters. 
As we use median values of posteriors($\omega_L$ and $\omega_H$), which have distributions in a 
certain level, and we use the constraint of $\omega_L \leq \omega_H$ in our calculation, 
it is reasonable to get some difference between $\omega_L$ and $\omega_H$. Since the typical errors  
of parameter inferences are 0.4 dex in the case, it is plausible to consider this is the main reason 
for the small diffferences. This guess can be supported by the fact that the slope of the distribution 
of these objects in Figure \ref{fig:covariance_function}(b) is well aligned to that of 'ideal' DRW 
model's one as shown by green dotted lines. 
Since the data we used for the analysis is not dense in time, 
it is very difficult to discuss the hypothesis of their intrinsic origin.  

\subsection{Issues for analyzing more massive data with our model}\label{subsec:calc_issues}

In our model the calculation time is the order of $O(N^{2})$, when $N$ is the number of observed data points. 
Compared to Kelly's OU mixture model, which needs time with the order of $O(MN^{2})$($M$ is the numbers of mixed OU processes), our 
model has an advantage in calculation time. However, it is still very massive for the data with numerous measurements, 
or with many observed objects. If we limit our analysis to those based on the stationary process, we may implement 
the methods based on phase-space description or Kalman filter for reducing the calculation time. 
Using the well-known time series analysis methods like ARMA (Auto-Regression Moving Average) model, as suggested by 
\citet{kelly2014, simm2016, kasliwal2017}, is among useful candidates for the massive data time series analysis in near future. 
It is, however, plausible to consider that some non-stationary processes also affect the optical variability of QSO/AGN. 
For analyzing time series data of QSO/AGN with a more complicated model, we should continue to strive for the effective 
algorithms using powerful computational equipment like many CPU cores and supportive database systems. 

\section{Conclusions} \label{sec:conclusions}
We develop the infinite mixture model of the OU process for describing the optical variability 
of QSOs/AGNs, which enables us to get the results within an appropriate time. 
The main reason for this faster calculation is our analytical deriving of the covariance function. 
It is based on the consideration for treating the variability as a stochastic process, and it enables us to 
get the parameters on their PSDs for their brightness variations. 

We apply our model to 67,507 objects extracted from SDSS Stripe82 photometric data with 
sufficient multiple measurements, and succeed in showing very high precision in selecting 
QSOs among variable objects based only on their variability, by investigating $\sim$10,000 
spectroscopically confirmed objects. 
This can be a good first step for enabling the classification and investigating the physical 
processes among various types of QSO/AGN using their variability. 

We find out that QSO/AGN variability can be well described in a certain regular manner in our model parameter 
space on their PSDs on QSO/AGN variability, with very few outliers. 
In general, the smaller the amplitude of variation, the shorter the damping time scale, which is consistent 
with the previous studies.   
The majority(95\%) of spectroscopically confirmed QSOs may be well described by DRW model. 
However, four percent of QSOs belong to the group with large differences between $\omega_L$ and $\omega_H$. 
The remaining one percent of QSOs are indistinguishable from stars in our model. 

Our model is more flexible than other previously suggested models based on a single OU process
(damped random walk), as we can describe their PSDs of variability with more accuracy.  
We eliminate the arbitrariness in the number of mixed OU processes from the original model,  
and our implementation enables calculations to finish within tolerable time.  
We show that this type of analysis is feasible in an era with more massive data, though more 
calculation speed is desirable. 

Based on our model, we try the separation of QSOs and stars based only on the variability, 
and succeed in identifying QSOs with $~$99.3\% completeness and 99.0\% purity. These numbers show 
our model can be used as the most effective method for selecting QSOs from a variable object catalog, 
and is superior to other models in terms of completeness. 
Other evaluations of the identification capabilities, such as accuracy and recall, show mostly the same 
levels as those for other competitive models, which are widely used and discussed for identifying QSOs 
based only on the variability. The main reason supporting our model's superiority in terms of QSO selection  
comes from our introduction of the mixture of OU processes, which provides more flexibility on the 
description of PSDs and enables us to separate rare-type QSOs/AGNs, which can not be well dexscribed by 
simple DRW model, from variable stars in the parameter space. 

\acknowledgments
We thank all the people who supported the research in many phase, especially to Prof. Yoshiyasu 
Tamura, Prof. Hidetoshi Konno, Prof. Shinichiro Koyama, Prof. Isao Shoji, Dr. Masaya Saito and Dr. Yuya Ariyoshi 
for their fruitful discussions in the preparation of the paper. 

This research was supported by the internship program of ``Data Scientist Training Network''
\footnote{http://datascientist.ism.ac.jp/}, a Commissioned Projects by 
The Ministry of Education, Culture, Sports, Science and Technology of Japan, 
and was also supported by JSPS KAKENHI Grant Number JP22012007. 

Funding for the SDSS and SDSS-II has been provided by the Alfred P. Sloan Foundation, 
the Participating Institutions, the National Science Foundation, the U.S. Department of Energy, 
the National Aeronautics and Space Administration, the Japanese Monbukagakusho, the Max Planck Society, 
and the Higher Education Funding Council for England. The SDSS Web Site is http://www.sdss.org/.

The SDSS is managed by the Astrophysical Research Consortium for the Participating Institutions. 
The Participating Institutions are the American Museum of Natural History, Astrophysical Institute Potsdam, 
University of Basel, University of Cambridge, Case Western Reserve University, University of Chicago, 
Drexel University, Fermilab, the Institute for Advanced Study, the Japan Participation Group, 
Johns Hopkins University, the Joint Institute for Nuclear Astrophysics, 
the Kavli Institute for Particle Astrophysics and Cosmology, the Korean Scientist Group, 
the Chinese Academy of Sciences (LAMOST), Los Alamos National Laboratory, 
the Max-Planck-Institute for Astronomy (MPIA), the Max-Planck-Institute for Astrophysics (MPA), 
New Mexico State University, Ohio State University, University of Pittsburgh, 
University of Portsmouth, Princeton University, the United States Naval Observatory, 
and the University of Washington.

Funding for SDSS-III has been provided by the Alfred P. Sloan Foundation, the Participating Institutions, 
the National Science Foundation, and the U.S. Department of Energy Office of Science. The SDSS-III web 
site is http://www.sdss3.org/.

SDSS-III is managed by the Astrophysical Research Consortium for the Participating Institutions of the 
SDSS-III Collaboration including the University of Arizona, the Brazilian Participation Group, 
Brookhaven National Laboratory, Carnegie Mellon University, University of Florida, 
the French Participation Group, the German Participation Group, Harvard University, 
the Instituto de Astrofisica de Canarias, the Michigan State/Notre Dame/JINA Participation Group, 
Johns Hopkins University, Lawrence Berkeley National Laboratory, Max Planck Institute for Astrophysics, 
Max Planck Institute for Extraterrestrial Physics, New Mexico State University, New York University, 
Ohio State University, Pennsylvania State University, University of Portsmouth, Princeton University, 
the Spanish Participation Group, University of Tokyo, University of Utah, Vanderbilt University, 
University of Virginia, University of Washington, and Yale University. 


%% for the paper.  Each keyword is check against the master list during
%% copy editing.  Individual instruments can be provided in parentheses,
%% after the keyword, but they are not verified.

\vspace{5mm}

\end{document}